\documentclass[journal=jctcce,manuscript=article]{achemso}

\usepackage{amsmath,amssymb}
\usepackage{graphics}
\usepackage{dcolumn}% Align table columns on decimal point
\usepackage{bm}% bold math
\usepackage[T1]{fontenc}
\usepackage[decmulti]{inputenc}
\usepackage{bbm}
\usepackage{txfonts}
\usepackage[bbgreekl]{mathbbol}
\usepackage{subfigure}
\usepackage[version=3]{mhchem}

\newcommand{\Aa}[0]{Aa}
\usepackage{colortbl}

\newcommand{\erf}{\ensuremath{\text{erf}}}

\newcommand{\lr}{\ensuremath{\text{lr}}}

\newcommand{\RSH}{\ensuremath{\text{RSH}}}

\newcommand{\RPA}{\ensuremath{\text{RPA}}}
\newcommand{\dRPA}{\ensuremath{\text{dRPA}}}
\newcommand{\RPAx}{\ensuremath{\text{RPAx}}}
\newcommand{\dRPAI}{\ensuremath{\text{dRPA-I}}}

\newcommand{\RPAxII}{\ensuremath{\text{RPAx-II}}}

\newcommand{\T}{\ensuremath{\text{T}}}

\newcommand{\bra}[1]{\ensuremath{\langle #1 \vert}}
\newcommand{\ket}[1]{\ensuremath{\vert #1  \rangle}}
\newcommand{\braket}[2]{\ensuremath{\langle  #1 \vert #2  \rangle}}

\renewcommand{\b}[1]{\ensuremath{\mathbf{#1}}}

\newcommand{\ua}{\ensuremath{\uparrow}}
\newcommand{\da}{\ensuremath{\downarrow}}

\newcommand{\onlinecite}[1]{\hspace{-1 ex}\nocite{#1}\citenum{#1}}

\newcommand{\highlight}[1]{{\color{black}\textrm{#1}}}
 
\author{J\'anos G. \'Angy\'an}
\email{angyan@crm2.uhp-nancy.fr}
\author{Ru-Fen Liu}\altaffiliation{Present address: Department of Chemistry and Biochemistry, M/C 9510, University of California, Santa Barbara, California 93106, USA} 
\affiliation{CRM2, Institut
  Jean Barriol, Nancy University and CNRS, 54506
  Vandoeuvre-l\`{e}s-Nancy, France }
\author{Julien Toulouse}
\email{julien.toulouse@upmc.fr}
\affiliation{Laboratoire de Chimie Th\'eorique, 
Universit\'e Pierre et Marie Curie and CNRS, 75005 Paris, France}
\author{Georg Jansen}
\email{georg.jansen@uni-due.de}
\affiliation{
Fakultät für Chemie, 
Universit\"at Duisburg-Essen, 
45117 Essen, Germany}

\date{\today}

\title[ACFDT correlation energies]{Correlation energy expressions from the adiabatic-connection fluctuation-dissipation theorem approach} 

\begin{document}
%%%%%%%%%%%%%%%%%%%%%%%%%%%%%%%%%%%%%%%%%%%%%%%%%%%%%%%%%%%%%%%
%% The manuscript does not need to include \maketitle, which is
%% executed automatically.  The document should begin with an
%% abstract, if appropriate.  
%% If one is given and should not be, 
%% the contents will be gobbled.
%%%%%%%%%%%%%%%%%%%%%%%%%%%%%%%%%%%%%%%%%%%%%%%%%%%%%%%%%%%%%%%
\begin{abstract}
We explore several random phase approximation (RPA) correlation energy variants within the adiabatic-connection fluctuation-dissipation theorem approach. These variants differ in the way the exchange interactions are treated. One of these variants, named dRPA-II, is original to this work and closely resembles the second-order screened exchange (SOSEX) method. We discuss and clarify the connections among different RPA formulations. We derive the spin-adapted forms of all the variants for closed-shell systems, and test them on a few atomic and molecular systems with and without range separation of the electron-electron interaction.
\end{abstract}

%%%%%%%%%%%%%%%%%%%%%%%%%%%%%%%%%%%%%%%%%%%%%%%%%%%%%%%%%%%%%%%
%% Start the main part of the manuscript here.
%%%%%%%%%%%%%%%%%%%%%%%%%%%%%%%%%%%%%%%%%%%%%%%%%%%%%%%%%%%%%%%

%%%%%%%%%%%%%%%%%%%%%%%%%%
\section{Introduction}
\label{sec:intro}
%%%%%%%%%%%%%%%%%%%%%%%%%%

There is a recent revival of interest in the random phase approximation (RPA) to obtain ground-state correlation energies of electronic systems~\cite{Yan:00,Furche:01b,Terakura:02,Miyake:02,Fuchs:02,Niquet:04,Fuchs:05,Furche:05,Dahlen:06,Marini:06,Jiang:07,Harl:08,Furche:08,Scuseria:08,Toulouse:09,Janesko:09,Janesko:09a,Janesko:09c,Ren:09,Lu:09,Harl:09,Nguyen:09,Gruneis:09,Ruzsinszky:10,Nguyen:10,Hellgren:10,Hesselmann:10,Paier:10,Harl:10,Ismail:10,Zhu:10,Eshuis:10,Toulouse:10a,Jansen:10,Lu:10,Hesselmann:11a,Ruzsinszky:11,Ren:11,Lotrich:11,Klopper:11,Hesselmann:11}. The RPA is considered as a promising first approximation to obtain non-perturbative, exact-exchange-compatible, post-Kohn-Sham correlation energy corrections in density-functional theory. In particular, the RPA is thought of as a remedy for the bad description of London dispersion forces by conventional local and semi-local density-functional approximations. However, it is widely admitted that while RPA is well adapted to long-range electron-electron interactions, for small interelectronic distances its performance is even poorer than that of semi-local density functionals~\cite{Kurth:99,Furche:01}. An efficient way to make an optimal use of RPA is to apply it in a range-separated approach~\cite{Toulouse:04a,Angyan:05}, where the short-range interactions are described by an exchange-correlation density functional, and long-range exchange and correlation are treated by Hartree-Fock (HF) and RPA, respectively. Computational schemes following these principles have been recently proposed and applied mainly to van der Waals complexes~\cite{Toulouse:09,Janesko:09,Janesko:09a,Paier:10,Zhu:10,Toulouse:10a,Toulouse:XX}.

Several formulations of RPA have been developed. Perhaps, the most well-known approach to RPA is the one based on the adiabatic-connection fluctuation-dissipation theorem (ACFDT)~\cite{Langreth:75,Langreth:77}. In this approach, the correlation energy expression involves integrations over both the frequency and the interaction strength, which can be performed either numerically or analytically. Obviously, an expression which has already been integrated analytically along at least one or both of these variables is more advantageous than the repeated use of numerical quadratures. If an analytical integration over the frequency is performed first, followed by a numerical integration over the interaction strength, one obtains an expression that is of the form of an interaction-strength-averaged two-particle density matrix contracted with the two-electron integrals. This is the \textit{adiabatic-connection formulation}. An analytical integration over the interaction strength followed by a numerical integration along the frequency leads to an expression involving the dynamic dielectric matrix. This is the \textit{dielectric-matrix formulation}. With a second analytical integration (either along the interaction strength starting from the adiabatic-connection expression, or along the frequency starting from the dielectric-matrix expression) both of these intermediate forms can be converted to a common expression, which consists in a sum of the shifts of electronic excitation energies when passing from an independent-particle to the RPA description of the excited states. This is the \textit{plasmon formulation}. The plasmon expression can be further converted to an equivalent expression involving coupled-cluster doubles (CCD) amplitudes calculated in the ring-diagram approximation~\cite{Scuseria:08}. This is the \textit{ring CCD formulation}. The relationship between the adiabatic-connection and ring CCD formulations of RPA has been recently discussed in Ref.~\onlinecite{Jansen:10}.

In this work, we study different variants of RPA within the adiabatic-connection formulation, which differ in the way the exchange interactions are handled. If the exchange interactions are neglected in the density matrix, we obtain the direct RPA (dRPA) approach (also called time-dependent Hartree), while inclusion of the non-local HF exchange response kernel leads to the RPAx approach (also called time-dependent Hartree-Fock, or full RPA). A third possibility, not discussed here, consists in including an exact exchange response kernel from a local exact exchange potential~\cite{Hesselmann:10}. If the dRPA density matrix is contracted with non-antisymmetrized two-electron integrals, one obtains what we call the dRPA-I variant, while if it is contracted with antisymmetrized two-electron integrals,
one obtains the dRPA-II variant. Similarly, if the RPAx density matrix is contracted with non-antisymmetrized two-electron integrals,  the RPAx-I variant is obtained, while if it is contracted with antisymmetrized two-electron integrals, one obtains the RPAx-II variant. The dRPA-I variant is just the commonly called ``RPA'' of the density-functional/material-science community. The dRPA-II variant, which is similar to the 
second-order screened exchange (SOSEX) expression introduced by Gr\"uneis {\it et al.}~\cite{Gruneis:09} in the ring CCD formulation, is original to this work. In contrast to SOSEX,
it involves higher-order screened exchange effects. The RPAx-II variant was first introduced by McLachlan and Ball~\cite{McLachlan:64}, but here we derive a new adiabatic-connection expression for it. Finally, the RPAx-I variant has been recently introduced by Toulouse {\it et al.}~\cite{Toulouse:09,Toulouse:10a}. When possible, for the case of dRPA-I and RPAx-II, we also compare with the equivalent plasmon formulation, and clarify the origin of the prefactor of $1/4$ in the plasmon formula of RPAx-II in place of the prefactor of $1/2$ appearing for dRPA-I. \highlight{We remind the reader that
in spite of the very different formulations, 
the  dRPA-I variant is the same as the direct 
ring-CCD  method, while the RPAx-II approach 
is identical to ring-CCD \cite{Jansen:10,Toulouse:XX}.}

For the sake of simplicity, we give all the expressions without range separation, but it is straightforward to generalize them for the case of range separation, as done in Ref.~\onlinecite{Toulouse:10a}. The paper is organized as follows. In Sec. II, we first provide an overview of the adiabatic-connection RPA correlation energy variants. In Sec. III, we review how the two-particle density matrix is obtained from the RPA polarization propagator. In Sec. IV, we derive the expressions of RPA correlation energy variants in spin-orbital basis. In Sec. V, we derive the corresponding spin-adapted expressions for closed-shell systems. In Sec. VI, we perform numerical comparisons of different variants on a few atomic and molecular systems with and without range separation. Finally, Sec. VII contains our conclusions. The analysis of the second-order limit in the electron-electron interaction of each variant is given in Appendix.

%%%%%%%%%%%%%%%%%%%%%%%%%%%%%%%%%%%%%%%%%%%
\section{Overview of RPA correlation energy variants in the adiabatic-connection formulation}
\label{sec:overview}
%%%%%%%%%%%%%%%%%%%%%%%%%%%%%%%%%%%%%%%%%%%

In the adiabatic-connection formalism, the correlation energy 
in a spin-orbital basis can be expressed as
\begin{equation}\label{eq:EcV}
  E_{c} = 
  \frac{1}{2} \int _0^1 \!\! d\alpha 
  \text{Tr}\left\{ \mathbb{V}\, 
  \mathbb{P}_{c,\alpha} \right\}
  = 
  \frac{1}{2} \int _0^1 \!\! d\alpha 
  \sum_{pq,rs} \braket{rq}{sp} 
  \left( \mathbb{P}_{c,\alpha} \right)_{pq,rs},
\end{equation}
where $\mathbb{V}_{sr,qp}=\braket{rq}{sp}$ are the two-electron 
integrals, $\mathbb{P}_{c,\alpha}$ is the correlation part of 
the two-particle density matrix at interaction strength 
$\alpha$, and $\text{Tr}$ denotes the trace (sum over the 
indices $rs$ and $pq$). Using the antisymmetry of 
$\mathbb{P}_{c,\alpha}$ 
with respect to the permutation of the indices $p$ and $s$, the 
correlation energy can also be expressed as
\begin{equation}\label{eq:EcW}
  E_{c} = 
  \frac{1}{4} \int _0^1 \!\! d\alpha 
  \text{Tr}\left\{ \mathbb{W}\, 
  \mathbb{P}_{c,\alpha} \right\}
  = 
  \frac{1}{4} \int _0^1 \!\! d\alpha 
  \sum_{pq,rs} \bra{rq}\ket{sp} 
  \left( \mathbb{P}_{c,\alpha} \right)_{pq,rs},
\end{equation}
where $\mathbb{W}_{sr,qp}=\bra{rq}\ket{sp}=
\braket{rq}{sp}-\braket{rq}{ps}$ are the antisymmetrized 
two-electron integrals. In RPA-type approximations, 
$\mathbb{P}_{c,\alpha}$ is obtained via the 
fluctuation-dissipation theorem
\begin{equation}\label{eq:fdtRPA}
  \mathbb{P}_{c,\alpha}^{\text{RPA}}  = 
  -\int_{-\infty }^{\infty }\! 
  \frac{d\omega}{2\pi i}\, e^{i\omega 0^+}  
  \left[ \Pi_{\alpha }^{\text{RPA}} (\omega) -   
  \Pi_{0} (\omega) \right],
\end{equation}
where $\Pi_{\alpha }^{\text{RPA}} (\omega)$ 
is the four-index matrix representation of dynamic polarization 
propagator at interaction strength $\alpha$ and frequency 
$\omega$, and $\Pi_{0}(\omega)$ is the corresponding 
non-interacting (Hartree-Fock or Kohn-Sham) polarization propagator. In the dRPA variant (or 
time-dependent Hartree) the polarization propagator is obtained 
from the response equation with the Hartree kernel 
$\mathbb{V}$
\begin{equation}\label{eq:DysonV}
  \Pi_{\alpha}^{\text{dRPA}}(\omega)^{-1} =   
  \Pi_{0}(\omega)^{-1} - \alpha \mathbb{V},
\end{equation}
whereas in the RPAx variant (or time-dependent Hartree-Fock) 
the polarization propagator is obtained using the Hartree-Fock 
kernel $\mathbb{W}$
\begin{equation}\label{eq:DysonW}
  \Pi_{\alpha}^{\text{RPAx}}(\omega)^{-1} =   
  \Pi_{0}(\omega)^{-1} - \alpha \mathbb{W}.
\end{equation}
The obtained dRPA and RPAx correlation density matrices 
$\mathbb{P}_{c,\alpha}^{\text{dRPA}}$ and 
$\mathbb{P}_{c,\alpha}^{\text{RPAx}}$ are completely 
expressed in the basis of occupied-virtual orbital products, 
i.e. $pq=ia$ or $ai$ and $rs=jb$ or $bj$ where $i,j$ refer to 
occupied orbitals and $a,b$ to virtual orbitals. Neither 
$\mathbb{P}_{c,\alpha}^{\text{dRPA}}$ nor 
$\mathbb{P}_{c,\alpha}^{\text{RPAx}}$ are properly 
antisymmetric. As a consequence, the two correlation energy 
expressions, \ref{eq:EcV} and~\ref{eq:EcW}, are no longer 
equivalent in dRPA or RPAx. This leads to at least four RPA 
variants for calculating correlation energies, denoted here by 
dRPA-I, dRPA-II, RPAx-I, and RPAx-II, depending whether the 
correlation density matrix is contracted with the 
non-antisymmetrized two-electron integrals $\mathbb{V}$ 
(variants I) or the antisymmetrized two-electron integrals 
$\mathbb{W}$ (variants II). 

The dRPA-I variant is obtained by inserting the dRPA correlation 
density matrix in \ref{eq:EcV},
\begin{align}\label{eq:DRPAI}
 E^{\text{dRPA-I}}_c&=
 \frac{1}{2}
 \int _0^1 \!\! d\alpha  
 \text{Tr}
 \left\{ 
 \mathbb{V}\,
 \mathbb{P}_{c,\alpha}^{\text{dRPA}} 
 \right\}.
\end{align}
This variant is commonly  called ``RPA'' in the 
density-functional/material-science community. It corresponds to 
the first RPA correlation energy approximation historically 
developed and is still widely used. Since the dRPA 
response equation involves the mere Hartree kernel, 
only the screening effect 
of the bare Coulomb interaction is taken into account in 
the polarization propagator and all exchange-correlation 
screening effects are neglected. The resulting correlation 
energies tend to be too strongly negative. At second order 
in the electron-electron interaction, the dRPA-I correlation 
energy does not reduce to the standard 
second-order M{\o}ller-Plesset (MP2) correlation energy, 
but instead to a ``direct MP2'' expression, i.e.\ without the MP2 exchange term~\cite{Furche:01b,Janesko:09b}.

The dRPA-II variant is obtained by contracting the dRPA 
correlation density matrix with the antisymmetrized two-electron 
integrals $\mathbb{W}$,
\begin{align}\label{eq:DRPAII}
 E^{\text{dRPA-II}}_c&=
 \frac{1}{2}
 \int _0^1 \!\! d\alpha \, 
 \text{Tr}
 \left\{ 
 \mathbb{W}\,
 \mathbb{P}^{\text{dRPA}}_{c,\alpha} 
 \right\},
\end{align}
which re-establishes the correct second-order MP2 limit. 
\ref{eq:EcW} could have suggested to use a factor of $1/4$ 
instead of $1/2$ in \ref{eq:DRPAII}, but in fact the 
correct MP2 limit is only recovered with the factor $1/2$. This 
variant can also be obtained from \ref{eq:DRPAI} by 
antisymmetrizing the correlation density matrix with respect to 
the permutation of $p$ and $s$: $
(\mathbb{P}_c^{\text{dRPA}})_{pq,rs} \to 
(\mathbb{P}_c^{\text{dRPA}})_{pq,rs}-
(\mathbb{P}_c^{\text{dRPA}})_{sq,rp}$. As far as we know, the 
dRPA-II variant has never been described before. 
It is similar to the second-order 
screened exchange (SOSEX) expression introduced by Gr\"uneis 
{\it et al.}~\cite{Gruneis:09} but the latter does not involve 
integration over the adiabatic connection and treats exchange
effects only at the lowest order of perturbation.

The RPAx-I variant is obtained by inserting the RPAx correlation 
density matrix in \ref{eq:EcV},
\begin{align}\label{eq:RPAXI}
 E^{\text{RPAx-I}}_c&=
 \frac{1}{2}
 \int _0^1 \!\! d\alpha  
 \,\text{Tr}
 \left\{ 
 \mathbb{V}\,
 \mathbb{P}^{\text{RPAx}}_{c,\alpha}  \right\},
\end{align}
and has been introduced recently by Toulouse 
{\it et al.}~\cite{Toulouse:09,Toulouse:10a}. In this variant, 
the exchange screening effects are taken into account in the 
polarization propagator. The matrix 
$\mathbb{P}^{\text{RPAx}}_{c,\alpha}$ is properly antisymmetric 
at first order, and therefore the RPAx-I correlation energy has 
the correct MP2 limit. At higher orders, however, 
$\mathbb{P}^{\text{RPAx}}_{c,\alpha}$ violates antisymmetry 
properties to some extent.

The RPAx-II variant is obtained by inserting the RPAx 
correlation density matrix in \ref{eq:EcW},
\begin{align}\label{eq:RPAXII}
 E^{\text{RPAx-II}}_c&=
 \frac{1}{4}
 \int _0^1 \!\! d\alpha  
 \,\text{Tr}
 \left\{ 
 \mathbb{W}\,
 \mathbb{P}^{\text{RPAx}}_{c,\alpha}  \right\},
\end{align}
which can also be obtained from \ref{eq:RPAXI} by 
antisymmetrizing the correlation density matrix: 
$(\mathbb{P}_c^{\text{RPAx}})_{pq,rs} \to (1/2) 
[(\mathbb{P}_c^{\text{RPAx}})_{pq,rs}-
(\mathbb{P}_c^{\text{RPAx}})_{sq,rp}]$, the factor $1/2$ being 
justified by the fact that $\mathbb{P}_c^{\text{RPAx}}$ is 
already approximately antisymmetric, in contrast to 
$\mathbb{P}_c^{\text{dRPA}}$. This variant was first introduced 
by McLachlan and Ball~\cite{McLachlan:64}. At second order, it 
properly reduces to MP2.

In the following, these four RPA correlation energy variants 
will be analyzed further and working expressions will be given.

%%%%%%%%%%%%%%%%%%%%%%%%%%%%%%%%%%%%%%%%%%%%%%%%%%%%%%%
\section{Two-particle density matrix from the 
polarization \\ propagator}
\label{sec:2pdm}
%%%%%%%%%%%%%%%%%%%%%%%%%%%%%%%%%%%%%%%%%%%%%%%%%%%%%%%

We first briefly review how to extract a two-particle density 
matrix from the RPA polarization propagator.
The non-interacting (Hartree-Fock or Kohn-Sham) polarization propagator 
$\Pi_0(\omega)$ writes
\begin{equation}
 \Pi_0(\omega) = 
 - (\Lambda_0 
 - \omega\,\Delta)^{-1},
\end{equation}
where $\Lambda_0$ and $\Delta$ are 
$2\times 2$ supermatrices
\begin{equation}
 \Lambda_0=
 \begin{pmatrix}
  \bm{\varepsilon}&   \b{0}  \\
  \b{0}& \bm{\varepsilon}     
 \end{pmatrix}               
 \qquad\quad \text{and}\qquad\quad 
 \Delta=
 \begin{pmatrix}
  \b{I}& \b{0}  \\
  \b{0}& -\b{I}     
 \end{pmatrix}.
\end{equation}
each block being of dimension $N_o N_v \times N_o N_v$, where 
$N_o$ and $N_v$ are the numbers of occupied and virtual 
orbitals, respectively. The diagonal matrix $\bm{\varepsilon}$ 
contains the independent one-particle excitation energies, 
$\varepsilon_{ia,jb}\!=\! (\varepsilon_a\!-\!
\varepsilon_i)\delta_{ij}\delta_{ab}$, and $\b{I}$ is the 
identity matrix. Similarly, the RPA polarization propagator at 
interaction strength $\alpha$ writes
\begin{equation}
  \Pi_\alpha^{\text{RPA}}(\omega) = 
  - (\Lambda_\alpha 
  - \omega\,\Delta)^{-1},
\end{equation}
where the supermatrix $\Lambda_\alpha$ is calculated 
with the Hartree kernel $\mathbb{V}$ in the case of dRPA,
\begin{equation}
  \Lambda^{\text{dRPA}}_\alpha = 
  \Lambda_0 + \alpha \mathbb{V},
\end{equation}
and with the Hartree-Fock kernel $\mathbb{W}$ 
in the case of RPAx,
\begin{equation}
  \Lambda^{\text{RPAx}}_\alpha = 
  \Lambda_0 + \alpha \mathbb{W}.
\end{equation}
From now on, we will consider real-valued orbitals. In this 
case, the Hartree kernel is made of four identical blocks,
\begin{equation}\label{eq:vmatrix}
 {\mathbb{V}}=
  \begin{pmatrix}
   \b{K}&   \b{K}  \\
   \b{K}& \b{K}     
  \end{pmatrix} ,
\end{equation}
where $K_{ia,jb} =\bra{ab} ij\rangle$ are 
non-antisymmetrized two-electron integrals. 
Similarly, the Hartree-Fock kernel writes
\begin{equation}\label{eq:wmatrix}
 {\mathbb{W}}=
 \begin{pmatrix}
 \b{A}^\prime&   \b{B}  \\
 \b{B}& \b{A}^\prime     
 \end{pmatrix} ,
\end{equation}
with the antisymmetrized two-electron integrals 
\begin{equation}\label{eq:Aprimedef}
 A^{\prime}_{ia,jb}= \bra{ib}\ket{aj}=
 \langle ib\vert aj\rangle -  
 \langle ib\vert ja\rangle =
 K_{ia,jb} - J_{ia,jb},
\end{equation}
and
\begin{equation}\label{eq:Bmatrixdef}
 B_{ia,jb}= \bra{ab}\ket{ij} =
 \langle ab\vert ij\rangle -  
 \langle ab\vert ji\rangle=
 K_{ia,jb} - K^\prime_{ia,jb}.
\end{equation}
Let us consider now the generalized non-hermitian RPA eigenvalue 
equation
\begin{align}\label{eq:rpaeigenpb}
  \Lambda_\alpha 
  \mathbb{C}_{\alpha,n} = \omega_{\alpha,n}
  \Delta\,
  \mathbb{C}_{\alpha,n},
\end{align}
whose solutions come in pairs: positive excitation energies 
$\omega_{\alpha,n}$ with eigenvectors 
$\mathbb{C}_{\alpha,n}=
\left( \b{x}_{\alpha,n}, \b{y}_{\alpha,n} \right)$ and negative 
excitation energies $\omega_{\alpha,-n}=-\omega_{\alpha,n}$ with 
eigenvectors $\mathbb{C}_{\alpha,-n}=
\left( \b{y}_{\alpha,n}, \b{x}_{\alpha,n} \right)$. The spectral 
representation of $\Pi_\alpha^{\text{RPA}}(\omega)$ 
then writes
\begin{align}\label{eq:rpaspectral}
  \Pi_{\alpha}^{\text{RPA}}(\omega) =\sum_n \left\{
  \frac{
  \mathbb{C}_{\alpha,n}^{~}
  \mathbb{C}_{\alpha,n}^\T}{\omega-\omega_{\alpha,n}+i0^+}
  -
  \frac{  
  \mathbb{C}_{\alpha,-n}^{~}
  \mathbb{C}_{\alpha,-n}^\T}{\omega-\omega_{\alpha,-n}-i0^+}
  \right\},
\end{align}
where the sum is over eigenvectors $n$ with positive excitation 
energies $\omega_{\alpha,n}>0$. The fluctuation-dissipation 
theorem [\ref{eq:fdtRPA}] leads to the supermatrix 
representation of the correlation density matrix 
$\mathbb{P}_{c,\alpha}^\RPA$ (using contour integration in 
the upper half of the complex plane)
\begin{align}\label{eq:PfromC} 
   \mathbb{P}_{c,\alpha}^\RPA = - 
   \int_{-\infty}^{\infty} 
   \frac{d\omega}{2\pi i} e^{i\omega0^+} 
   [ \Pi_\alpha^\RPA (\omega) - 
     \Pi_0 (\omega)] =
   \sum_n
    \left\{ 
         \mathbb{C}_{\alpha,-n}^{~}
         \mathbb{C}_{\alpha,-n}^{\T}
         -
         \mathbb{C}_{0,-n}^{~}
         \mathbb{C}_{0,-n}^{\T}
 \right\},
\end{align}
with the non-interacting eigenvectors 
$\mathbb{C}_{0,-n}=\left( \b{y}_{0,n}, \b{x}_{0,n} \right)$ 
with $\b{y}_{0,n}=\b{0}$ and $\b{x}_{0,n}=\b{1}_{n}$ (where 
$\b{1}_{n}$ is the vector whose $n^\text{th}$ component is 
$1$ and all other components are zero). The explicit 
supermatrix expression of the RPA correlation 
density matrix is thus
\begin{align}\label{eq:PfromXY} 
\mathbb{P}_{c,\alpha}^\RPA = 
   \left( 
   \begin{array}{cc} 
   \b{Y}_{\alpha}^{~} \b{Y}_{\alpha}^{\T} & 
   \b{Y}_{\alpha}^{~} \b{X}_{\alpha}^{\T} \\ 
   \b{X}_{\alpha}^{~} \b{Y}_{\alpha}^{\T} & 
   \b{X}_{\alpha}^{~} \b{X}_{\alpha}^{\T} \\ 
   \end{array} 
   \right) - \left( 
   \begin{array}{cc} 
   \b{0} & \b{0}\\  
   \b{0} & \b{I}\\ 
   \end{array} 
   \right),
\end{align}
where $\b{X}_{\alpha}$ and $\b{Y}_{\alpha}$ are the matrices whose columns contain the eigenvectors $\b{x}_{\alpha,n}$ 
and $\b{y}_{\alpha,n}$.
The dRPA and RPAx correlation density matrices have the same 
form in terms of the eigenvector matrices $\b{X}_{\alpha}$ and 
$\b{Y}_{\alpha}$, although the eigenvectors are of course 
different for dRPA and RPAx.

%%%%%%%%%%%%%%%%%%%%%%%%%%%%%%%%%%%%%%%%%%%%%%%%%%%%%%%%%%%
\section{Correlation energy expressions in spin-orbital basis}
\label{sec:spinorb}
%%%%%%%%%%%%%%%%%%%%%%%%%%%%%%%%%%%%%%%%%%%%%%%%%%%%%%%%%%%

We give here the expressions in a spin-orbital basis for 
calculating the different RPA correlation energy variants. We 
first consider the dRPA-I and RPAx-II variants which have 
similar expressions. In both cases the integration over the 
adiabatic connection can be done analytically, leading to 
plasmon formulae. We then examine the dRPA-II and RPAx-I 
variants. They have in common that they are mixing the non-antisymmetrized
integrals $\mathbb{V}$ and the antisymmetrized 
integrals $\mathbb{W}$, which makes it impossible to do the 
integration over the adiabatic connection analytically. Although 
the dRPA-I variant is well-documented in the literature after 
the work of Furche and 
coworkers~\cite{Furche:01b,Furche:08,Eshuis:10}, the review that 
we give here is useful to define 
our notations and for comparisons with other variants. The RPAx-
I variant has been discussed in detail in the context of range 
separation by Toulouse 
{\it et al.}~\cite{Toulouse:09,Toulouse:10a,Zhu:10}. 
The RPAx-II variant is much less documented and the 
dRPA-II is new, so most of the expressions that we give for them 
are original to this work.

%%%%%%%%%%%%%%%%%%%%%%%%%%%%%%%%%%%%%%%%%%%%%%%%%%%%%%%
\subsection{dRPA-I correlation energy}
\label{sec:drpaI}
%%%%%%%%%%%%%%%%%%%%%%%%%%%%%%%%%%%%%%%%%%%%%%%%%%%%%%%

There are several equivalent expressions for the dRPA-I 
correlation energy.

\subsubsection{Adiabatic-connection formula}
\label{sec:ac-drpaI}
The dRPA-I correlation energy of \ref{eq:DRPAI} can be 
expressed with the eigenvectors of the dRPA polarization 
propagator according to the general prescription to form the 
correlation density matrix, \ref{eq:PfromC},
\begin{align}\label{eq:dRPAI}
 E_c^{\text{dRPA-I}}&=\frac{1}{2}
 \int _0^1 \!\! d\alpha \sum_n\text{Tr}
         \left\{
         \mathbb{V}\,
         \mathbb{C}_{\alpha,-n}^{~}
         \mathbb{C}_{\alpha,-n}^{\T} 
        -
         \mathbb{V}\,
         \mathbb{C}_{0,-n}^{~}
         \mathbb{C}_{0,-n}^{\T} 
         \right\},
\end{align}
or, using the explicit expressions in terms of the block matrix 
components [\ref{eq:vmatrix} and \ref{eq:PfromXY}],
\begin{align}
  E_c^{\text{dRPA-I}}=&
  \frac{1}{2} 
  \int _0^1 d\alpha  \,
  \text{tr}
  \left\{
  \left[
  \left(
  \b{X}_{\alpha}+ 
  \b{Y}_{\alpha} \right)
  \left(
  \b{X}_{\alpha} +
  \b{Y}_{\alpha} \right)^\T-
  \b{I}
  \right]
  \b{K}
  \right\}, 
 \end{align}
where $\text{tr}$ refers to the trace now applied to the block 
matrices (which are half the size of the supermatrices). As 
shown by Furche~\cite{Furche:01b}, one does not need to 
calculate explicitly the eigenvector matrices
$\b{X}_{\alpha}$ and $\b{Y}_{\alpha}$ to get the 
correlation energy; it is sufficient to form the 
matrix
\begin{equation}\label{eq:Qa}
  \b{Q}_\alpha = 
  \left(\b{X}_{\alpha} + 
  \b{Y}_{\alpha}\right) \left(\b{X}_{\alpha} + 
  \b{Y}_{\alpha}\right)^\T,
\end{equation}
which can be obtained directly from the matrices involved 
in the RPA response equation. In the case of dRPA, 
it simply reads
\begin{align}\label{eq:QdRPA}
  \b{Q}^\text{dRPA}_\alpha = 
  \bm{\varepsilon}^{1/2} 
  \left(\b{M}^\text{dRPA}_\alpha\right)^{-1/2}
  \bm{\varepsilon}^{1/2},
\end{align}
with
\begin{equation}\label{eq:MdRPA}
  \b{M}^\text{dRPA}_\alpha = 
  \bm{\varepsilon}^{1/2}\,
 (\bm{\varepsilon}+2 \,\alpha \,\b{K})\, 
  \bm{\varepsilon}^{1/2}.
\end{equation}
The adiabatic-connection formula for the dRPA-I correlation 
energy is then finally
\begin{equation}
\label{eq:Ec_singlebar}
  E_c^{\text{dRPA-I}}=
  \frac{1}{2}\int_0^1 d\alpha \, \text{tr}
  \left\{\left[\b{Q}_{\alpha }^{\text{dRPA}} -
  \b{I} \right]\,\b{K}\right\}.
\end{equation} 
In previous papers, this equation was written with the matrix 
$\b{P}_{\alpha}^{\text{dRPA}}=
\b{Q}_{\alpha }^{\text{dRPA}} -\b{I}$.

\subsubsection{Plasmon formula}
\label{sec:plasmon-drpaI}
The plasmon formula for the dRPA-I correlation energy is found 
by starting from an equivalent form of \ref{eq:dRPAI},
\begin{align}\label{eq:ecorrIc}
 E_c^{\text{dRPA-I}}&=\frac{1}{2}
 \int _0^1 \!\! d\alpha \sum_n\text{Tr}
         \left\{
         \mathbb{C}_{\alpha,-n}^{\T}
         \mathbb{V}\,
         \mathbb{C}_{\alpha,-n}^{~}
        -
         \mathbb{C}_{0,-n}^{\T}
         \mathbb{V}\,
         \mathbb{C}_{0,-n}^{~}
         \right\},
\end{align}
obtained by a cyclic permutation of the matrices in the trace. 
Since the positive excitation energies can be written 
as~\cite{McLachlan:64,Furche:08} 
\begin{equation}
  \omega_{\alpha,n}^\dRPA = 
  \mathbb{C}_{\alpha,-n}^\T
  \Lambda_\alpha^{\text{dRPA}}
  \mathbb{C}_{\alpha,-n}^{~},
\end{equation}
the derivative of $\omega_{\alpha,n}$ with respect to $\alpha$ 
gives
\begin{equation}\label{eq:derivomegan}
  \frac{d\omega_{\alpha,n}^\dRPA}{d \alpha} = 
  \mathbb{C}_{\alpha,-n}^{\T} 
  \frac{d\Lambda^{\text{dRPA}}_{\alpha}}{d \alpha}
         \mathbb{C}_{\alpha,-n}^{~}
 =
         \mathbb{C}_{\alpha,-n}^{\T} 
         \mathbb{V}
         \mathbb{C}_{\alpha,-n}^{~},
\end{equation}
which allows one to perform the integral over $\alpha$ in 
\ref{eq:ecorrIc} analytically, leading to the plasmon 
formula
\begin{align}\label{eq:ecorr_plasmon}
 E_c^{\text{dRPA-I}}&=\frac{1}{2} \sum_n
 \left(
 \omega_{1,n}^{\dRPA} - \omega_{0,n}
       -
         \mathbb{C}_{0,-n}^{\T} 
         \mathbb{V}
         \mathbb{C}_{0,-n}^{~}
         \right)
         \nonumber \\ &
         =\frac{1}{2}
         \sum_n
 \left( \omega_{1,n}^{\text{dRPA}} - \omega_{n}^{\text{dTDA}}
 \right), 
\end{align}
where $\sum_n \omega_{n}^{\text{dTDA}} = \sum_n \mathbb{C}_{0,-n}^{\T} 
\Lambda_1^{\text{dRPA}} \mathbb{C}_{0,-n}^{~} = \sum_n
\omega_{0,n}+\mathbb{C}_{0,-n}^{\T} \mathbb{V} 
\mathbb{C}_{0,-n}^{~}$ is the sum of the (positive) excitation energies in the direct 
Tamm-Dancoff approximation (dTDA). The sum of the dTDA 
excitation energies can also be expressed as $\sum_n \omega_{n}^{\text{dTDA}} 
= \text{tr}\{\bm{\varepsilon}+\b{K}\}$.

\subsubsection{Alternative plasmon formula}
\label{sec:altplasmon-drpaI}
An alternative form of the plasmon formula can be found by 
rewriting \ref{eq:ecorr_plasmon} as
\begin{align}\label{eq:ecorrIalt}
 E_c^{\text{dRPA-I}}&=\frac{1}{2}
\sum_n\text{Tr}
         \left\{
         \Lambda_1^{\text{dRPA}}\,
         \mathbb{C}_{1,-n}^{~}
         \mathbb{C}_{1,-n}^{\T} 
        -
         \Lambda_1^{\text{dRPA}}\,
         \mathbb{C}_{0,-n}^{~}
         \mathbb{C}_{0,-n}^{\T} 
         \right\},
\end{align}
where the cyclic invariance of the trace has again been used. 
Using then \ref{eq:PfromXY} and recalling that the 
diagonal blocks of $\Lambda_1^{\text{dRPA}}$ are 
$\bm{\varepsilon}+\b{K}$ and the off-diagonal blocks are 
$\b{K}$, the correlation energy becomes
\begin{align}
  E_c^{\text{dRPA-I}}=&
  \frac{1}{2}  
  \text{tr}
  \left\{
  \left[
  \b{Y}_{1}^{~} 
  \b{Y}_{1}^\T +
  \b{X}_{1}^{~}
  \b{X}_{1}^\T-
  \b{I}
  \right]
  \left(\bm{\varepsilon} +\b{K}\right)+
\left[
  \b{Y}_{1}^{~}
  \b{X}_{1}^\T+
  \b{X}_{1}^{~}
  \b{Y}_{1}^\T
  \right]\b{K}
  \right\}.
\end{align}
Introducing now the inverse of the $\b{Q}_{\alpha}$ 
matrix~\cite{Furche:01},
\begin{equation}\label{eq:Qainv}
   \b{Q}_\alpha^{-1} = %\sum_n 
   \left(\b{X}_{\alpha} - 
   \b{Y}_{\alpha}\right) \left(\b{X}_{\alpha} - 
   \b{Y}_{\alpha}\right)^\T,
\end{equation}
which in the case of dRPA can be written as
\begin{equation}\label{eq:QinvdRPA}
  \left(\b{Q}^\text{dRPA}_\alpha\right)^{-1}=
  \bm{\varepsilon}^{-1/2} 
  \left(\b{M}^\text{dRPA}_\alpha\right)^{1/2}
  \bm{\varepsilon}^{-1/2},
\end{equation}
the correlation energy can be expressed as
\begin{align}\label{eq:dRPAnointeg}
 E_c^{\text{dRPA-I}}
    = \frac{1}{2}\text{tr}
    \biggl\{\left[ \tfrac{1}{2} 
    \left(\b{Q}_1^{\text{dRPA}}+
    \left(\b{Q}_1^{\text{dRPA}}\right)^{-1}\right) -
    \b{I}\right]\left(\bm{\varepsilon}+\b{K} \right)
 +  \tfrac{1}{2} \left(\b{Q}_1^{\text{dRPA}}-   
    \left(\b{Q}_1^{\text{dRPA}}\right)^{-1}\right)\b{K} 
    \biggl\},
\end{align}
or, equivalently,
\begin{align}\label{eq:dRPAnointeg2}
 E_c^{\text{dRPA-I}}
    = \frac{1}{2}\text{tr}\biggl\{
 \left[\b{Q}_1^{\text{dRPA}}-\b{I}\right]\b{K}
 +
 \tfrac{1}{2} \left[ \b{Q}_1^{\text{dRPA}}+
 \left(\b{Q}_1^{\text{dRPA}}\right)^{-1}-2\b{I}\right]
 \bm{\varepsilon}
 \biggr\},
\end{align}
or, rearranged in a different way
\begin{align}
   E_c^{\text{dRPA-I}}
          = \frac{1}{2}\text{tr}\biggl\{
   \tfrac{1}{2}\b{Q}_1^{\text{dRPA}}
        \left(\bm{\varepsilon}+2\b{K}\right)
 +
   \tfrac{1}{2}\left(\b{Q}_1^{\text{dRPA}}\right)^{-1}
        \,\bm{\varepsilon} - 
         \, \left(\bm{\varepsilon} +\b{K}\right)
 \biggr\}.
\end{align}
Using the expressions of $\b{Q}_1^{\text{dRPA}}$ 
[\ref{eq:QdRPA}], $(\b{Q}_1^{\text{dRPA}})^{-1}$ 
[\ref{eq:QinvdRPA}], and $\b{M}_1^{\text{dRPA}}$ 
[\ref{eq:MdRPA}], and the cyclic invariance of the trace, 
we finally arrive at the alternative form of the plasmon formula 
for the dRPA-I correlation energy
\begin{align}\label{eq:dRPAI-apf}
   E_c^{\text{dRPA-I}}
          = \frac{1}{2}\text{tr}\biggl\{
 \left(\b{M}_1^{\text{dRPA}}\right)^{1/2}
 - \left(\bm{\varepsilon}+\b{K}\right)
 \biggr\}.
\end{align}
Recently, \ref{eq:dRPAI-apf} 
have been used by 
Eshuis~{\it et al.}~\cite{Eshuis:10} as the starting point for 
developing a computationally efficient algorithm for calculating 
the dRPA-I correlation energy. Note that 
expression~\ref{eq:dRPAI-apf} could 
have also been found by noting that the eigenvalues of 
$\b{M}_1^{\text{dRPA}}$ are $(\omega_{1,n}^{\text{dRPA}})^2$ and 
thus $\sum_n \omega_{1,n}^{\text{dRPA}} = 
\text{tr}\{ (\b{M}_1^{\text{dRPA}})^{1/2} \}$. However, working 
with $\b{Q}_\alpha^{-1}$ will be useful for the other variants. 
Also, comparison of \ref{eq:Ec_singlebar} and 
\ref{eq:dRPAnointeg2} provides us with a decomposition of 
the correlation energy into kinetic and potential contributions, 
$E_c^\dRPAI = T_c^\dRPAI + U_c^\dRPAI$. Indeed, the potential 
correlation energy is just the value of the integrand in 
\ref{eq:Ec_singlebar} at $\alpha=1$, i.e. 
\begin{equation}
   U_c^\dRPAI = 
   \frac{1}{2} \text{tr}\left\{ 
   \left[ \b{Q}_1^{\dRPA} -\b{I} \right] \b{K} \right\},
\end{equation}
and thus, by subtraction, according to
\ref{eq:dRPAnointeg2}, 
the kinetic correlation energy is 
\begin{equation}\label{eq:TcdRPAI}
   T_c^\dRPAI = 
   \frac{1}{4} \text{tr}\left\{ 
   \left[ \b{Q}_1^{\dRPA} + 
   \left( \b{Q}_1^{\dRPA}\right)^{-1}-2\b{I} \right]
   \bm{\varepsilon}\right\}.
\end{equation}
In the limit of a system with orbitals that are all degenerate, 
i.e.\ with static correlation only, then 
$\bm{\varepsilon}=\b{0}$ and the kinetic correlation energy 
vanishes as it should. This is in agreement with the statement that 
dRPA-I correctly describes left-right static correlation in 
bond dissociations~\cite{Fuchs:05,Henderson:10}.

%%%%%%%%%%%%%%%%%%%%%%%%%%%%%%%%%%%%%%%%%%%%%%%%%%%%%%%
\subsection{RPAx-II correlation energy}
\label{sec:drpaxII}
%%%%%%%%%%%%%%%%%%%%%%%%%%%%%%%%%%%%%%%%%%%%%%%%%%%%%%%

We now derive several equivalent RPAx-II correlation energy 
expressions by proceeding in an analogous way to the case of 
dRPA-I.

\subsubsection{Adiabatic-connection formula}
\label{sec:ac-rpaxII}

The RPAx-II correlation energy of \ref{eq:RPAXII} can be 
written in terms of the eigenvectors of the RPAx polarization 
propagator
\begin{align}\label{eq:EcRPAxII}
 E_c^{\text{RPAx-II}}=\frac{1}{4}
 \int _0^1 \!\! d\alpha \sum_n\text{Tr}
         \left\{
         \mathbb{W}\,
         \mathbb{C}_{\alpha,-n}^{~}
         \mathbb{C}_{\alpha,-n}^{\T} 
        -
         \mathbb{W}\,
         \mathbb{C}_{0,-n}^{~}
         \mathbb{C}_{0,-n}^{\T} 
         \right\},
\end{align}
or, using the block structure of $\mathbb{W}$ 
[\ref{eq:wmatrix}], 
\begin{align}
 E_c^{\text{RPAx-II}}=
  \frac{1}{4} 
  \int _0^1 d\alpha  \,
  %\sum_n
  \text{tr}
  \left\{\left(
  \b{Y}_{\alpha}^{~} 
  \b{Y}_{\alpha}^\T+
  \b{X}_{\alpha}^{~}
  \b{X}_{\alpha}^\T-
  \b{I}
  \right)
  \b{A'}
  +
  \left(
  \b{Y}_{\alpha}^{~}
  \b{X}_{\alpha}^\T+
  \b{X}_{\alpha}^{~}
  \b{Y}_{\alpha}^\T
  \right)\b{B}
  \right\}.
\end{align}
Using the matrix $\b{Q}_\alpha$ which in the case of RPAx 
is given by
\begin{align}\label{eq:QRPAx}
  \b{Q}^\text{RPAx}_\alpha = 
  \left(\bm{\varepsilon}+\alpha\b{A}^\prime-\alpha\b{B}
  \right)^{1/2}
  \left(\b{M}_\alpha^\RPAx \right)^{-1/2}
  \left(\bm{\varepsilon}+\alpha\b{A}^\prime-
  \alpha\b{B}\right)^{1/2},
\end{align}
with 
\begin{equation}\label{eq:MRPAx}
   \b{M}^\text{RPAx}_\alpha = 
   \left(\bm{\varepsilon}+\alpha\b{A}^\prime-
   \alpha\b{B}\right)^{1/2} 
   \left(\bm{\varepsilon}+\alpha\b{A}^\prime+
   \alpha\b{B}\right) 
   \left(\bm{\varepsilon}+\alpha\b{A}^\prime-
   \alpha\b{B}\right)^{1/2},
\end{equation}
and the inverse $\b{Q}_\alpha^{-1}$
\begin{align}\label{eq:QRPAxinv}
    \left(\b{Q}^\text{RPAx}_\alpha\right)^{-1} =
    \left(\bm{\varepsilon}+\alpha\b{A}^\prime-
    \alpha\b{B}\right)^{-1/2}
    \left(\b{M}_\alpha^\RPAx \right)^{1/2}
    \left(\bm{\varepsilon}+\alpha\b{A}^\prime-
    \alpha\b{B}\right)^{-1/2},
\end{align}
we arrive at the adiabatic-connection formula for the 
RPAx-II correlation energy
\begin{align}\label{eq:ecorr_acfdt2b}
    E_c^{\text{RPAx-II}}=
    \frac{1}{4}\int _0^1 d\alpha \,  \text{tr}\left\{ 
    \tfrac{1}{2}\b{Q}_{\alpha}^{\text{RPAx}} 
    \left(\b{A}'+\b{B}\right)+
    \tfrac{1}{2}\left(\b{Q}_{\alpha }^{\text{RPAx}}\right)^{-1}
    \left(\b{A}'-\b{B}\right)-\b{A}'\right\}.
\end{align}

Since $\b{Q}_\alpha=\b{I}+\b{P}_{\alpha}$, if $\b{P}_{\alpha}$ 
is small, we can consider the approximation
$\b{Q}_\alpha^{-1} = (\b{I}+\b{P}_{\alpha})^{-1}
\approx\b{I}-\b{P}_{\alpha}=2\b{I}-\b{Q}_{\alpha}$, 
which leads to the following approximation for the 
RPAx-II correlation energy
\begin{align}\label{eq:RPAxIIapprox}
    E_c^{\text{RPAx-IIa}} &=
    \frac{1}{4}\int _0^1 d\alpha \, \text{tr}\left\{ 
    \tfrac{1}{2}\b{Q}_{\alpha }^{\text{RPAx}} 
    \left(\b{A}'+\b{B}\right)+
    \tfrac{1}{2}\left(2 \b{I} - 
    \b{Q}_{\alpha }^{\text{RPAx}}\right)
    \left(\b{A}'-\b{B}\right)-\b{A}'\right\}
\nonumber\\
 &=
  \frac{1}{4}\int _0^1d\alpha  \,
  \text{tr}\left\{\left[ \b{Q}_{\alpha}^{\text{RPAx}} -
  \b{I}\right]\,\b{B}\right\}.
\end{align}
So, we have the interesting result that this approximate 
correlation energy expression is analogous to the dRPA-I 
correlation energy expression of \ref{eq:Ec_singlebar}, 
the only differences being that the matrix 
$\b{Q}_{\alpha}$ is now obtained from the RPAx response 
equation and that it is contracted with the antisymmetrized 
two-electron integrals $\b{B}$, along with the corresponding 
change of the prefactor from 1/2 to 1/4.

\subsubsection{Plasmon formula}
\label{sec:plasmon-rpaxII}

As in the case of dRPA-I, the plasmon formula for the RPAx-II 
correlation energy is found by taking profit of the cyclic 
invariance of the trace to rewrite \ref{eq:EcRPAxII} as
\begin{align}\label{eq:ecorrspecral}
 E_c^{\text{RPAx-II}}&=\frac{1}{4}
 \int _0^1 \!\! d\alpha \sum_n\text{Tr}
         \left\{ 
         \mathbb{C}_{\alpha,-n}^{\T} 
         \mathbb{W}
         \mathbb{C}_{\alpha,-n}^{~}
        -
         \mathbb{C}_{0,-n}^{\T} 
         \mathbb{W}
         \mathbb{C}_{0,-n}^{~}
         \right\},
\end{align}
and then using $d\omega_{\alpha,n}^\RPAx/d\alpha =
\mathbb{C}_{\alpha,-n}^{\T} 
(d\Lambda^\RPAx_{\alpha}/d\alpha) 
\mathbb{C}_{\alpha,-n}^{~} =
\mathbb{C}_{\alpha,-n}^{\T}\mathbb{W} 
\mathbb{C}_{\alpha,-n}^{~}$ to integrate analytically over 
$\alpha$
\begin{align}\label{eq:ecorr_plasmonII}
 E_c^{\text{RPAx-II}}&=\frac{1}{4} \sum_n
 \left(
 \omega_{1,n}^{\RPAx} - \omega_{0,n}
       -
         \mathbb{C}_{0,-n}^{\T} 
         \mathbb{W}
         \mathbb{C}_{0,-n}^{~}
         \right)
         \nonumber \\ &
     = \frac{1}{4}
         \sum_n
 \left( \omega_{1,n}^{\text{RPAx}} - 
 \omega_{n}^{\text{TDAx}}
 \right),
\end{align}
where $\sum_n \omega_{n}^{\text{TDAx}} = \sum_n
\mathbb{C}_{0,-n}^{\T} \Lambda_1^{\text{RPAx}} 
\mathbb{C}_{0,-n}^{~} = \sum_n \omega_{0,n}+
\mathbb{C}_{0,-n}^{\T} \mathbb{W} 
\mathbb{C}_{0,-n}^{~}$ is the sum of the (positive) excitation energies in the 
Tamm-Dancoff approximation with exchange (TDAx) or configuration 
interaction singles (CIS). The sum of the TDAx 
excitation energies can also be expressed as $\sum_n \omega_{n}^{\text{TDAx}} = 
\text{tr} \left\{ \bm{\varepsilon} + \b{A}^\prime\right\}$. 
This plasmon formula was first presented by McLachlan and 
Ball~\cite{McLachlan:64}. The presence of a factor of $1/4$ in 
\ref{eq:ecorr_plasmonII} and not a factor of $1/2$ like in 
\ref{eq:ecorr_plasmon} has been debated in the 
literature~\cite{Oddershede:78}. The present exposition makes 
it clear that this factor of $1/4$ is due to the use of the 
antisymmetrized two-electron integrals $\mathbb{W}$. 

\subsubsection{Alternative plasmon formula}
\label{sec:altplasmon-rpaxII}

As in the case of dRPA-I, the alternative plasmon formula is 
found by rewriting \ref{eq:ecorr_plasmonII} as
\begin{align}\label{eq:ecorrI}
 E_c^{\text{RPAx-II}}&=\frac{1}{4}
 \sum_n\text{Tr}
         \left\{
         \Lambda_1^{\text{RPAx}}\,
         \mathbb{C}_{1,-n}^{~}
         \mathbb{C}_{1,-n}^{\T} 
        -
         \Lambda_1^{\text{RPAx}}\,
         \mathbb{C}_{0,-n}^{~}
         \mathbb{C}_{0,-n}^{\T} 
         \right\},
\end{align}
and inserting the diagonal blocks of 
$\Lambda_1^{\text{RPAx}}$ which are $\bm{\varepsilon}+\b{A}^\prime$ and the off-diagonal blocks 
which are $\b{B}$, 
\begin{align}\label{eq:RPAxIIplasmon}
    E_c^{\text{RPAx-II}}=
    \frac{1}{4} \text{tr}\left\{ 
    \tfrac{1}{2}\b{Q}_{1}^{\text{RPAx}} 
    \left(\bm{\varepsilon}+\b{A}^\prime+\b{B}\right)+
    \tfrac{1}{2}\left(\b{Q}_{1}^{\text{RPAx}}\right)^{-1}
    \left(\bm{\varepsilon}+\b{A}^{\prime}-\b{B}\right)-
    (\bm{\varepsilon}+\b{A}^\prime)\right\}.
\end{align}
Using the expressions of $\b{Q}_{1}^{\text{RPAx}}$ 
[\ref{eq:QRPAx}], 
$\left(\b{Q}_{1}^{\text{RPAx}}\right)^{-1}$ 
[\ref{eq:QRPAxinv}], and $\b{M}_1^{\text{RPAx}}$ 
[\ref{eq:MRPAx}], and the cyclic invariance of the trace, 
we arrive at the alternative plasmon formula for the RPAx-II 
correlation energy
\begin{align}
   E_c^{\text{RPAx-II}}
          = \frac{1}{4}\text{tr}\biggl\{
 \left(\b{M}_1^{\text{RPAx}}\right)^{1/2}
 - \left(\bm{\varepsilon}+ \b{A}^\prime\right)
 \biggr\}.
\end{align}
Finally, just as for dRPA-I, comparison of 
\ref{eq:ecorr_acfdt2b} and \ref{eq:RPAxIIplasmon} 
provides us with a decomposition of the correlation energy into 
the potential energy contribution to the correlation energy
\begin{equation}
  U_c^\RPAxII =\frac{1}{4} \text{tr}\left\{ 
    \tfrac{1}{2}\b{Q}_{1}^{\text{RPAx}} 
    \left(\b{A}^\prime+\b{B}\right)+
    \tfrac{1}{2}\left(\b{Q}_{1}^{\text{RPAx}}\right)^{-1}
    \left(\b{A}^{\prime}-\b{B}\right)-\b{A}^\prime\right\},
\end{equation}
and the kinetic correlation energy 
\begin{equation}\label{eq:TcRPAx}
   T_c^\RPAxII = \frac{1}{8} 
   \text{tr}\left\{ \left[ \b{Q}_1^{\RPAx} + 
   \left( \b{Q}_1^{\RPAx}\right)^{-1}-2\b{I} \right] 
   \bm{\varepsilon}\right\}.
\end{equation}
The RPAx-II kinetic correlation energy vanishes in the 
limit where $\bm{\varepsilon}=\b{0}$ as for dRPA-I.

%%%%%%%%%%%%%%%%%%%%%%%%%%%%%%%%%%%%%%%%%%%%%%%%%%%%%
\subsection{dRPA-II correlation energy}
\label{sec:drpaII}
%%%%%%%%%%%%%%%%%%%%%%%%%%%%%%%%%%%%%%%%%%%%%%%%%%%%%

The dRPA-II correlation energy of \ref{eq:DRPAII}
writes in terms of the eigenvectors of the dRPA polarization 
propagator
\begin{align}\label{eq:dRPAII}
   E_c^{\text{dRPA-II}}&=\frac{1}{2}
         \int _0^1 \!\! d\alpha \sum_n\text{Tr}
         \left\{
         \mathbb{W}\,
         \mathbb{C}_{\alpha,-n}^{~}
         \mathbb{C}_{\alpha,-n}^{\T} 
        -
         \mathbb{W}\,
         \mathbb{C}_{0,-n}^{~}
         \mathbb{C}_{0,-n}^{\T} 
         \right\},
\end{align}
leading to 
\begin{align}\label{eq:RPASX_dm}
  E_c^{\text{dRPA-II}} =
  \frac{1}{2}\int_0^1 d\alpha \, 
  \text{tr}\left\{ 
  \tfrac{1}{2}
  \b{Q}_{\alpha}^{\text{dRPA}} 
  \left(\b{A}'+\b{B}\right) 
  +
  \tfrac{1}{2}
  \left(\b{Q}_{\alpha }^{\text{dRPA}}\right)^{-1} 
  \left(\b{A}'-\b{B}\right)-\b{A}'
  \right\}.
\end{align}
Equation~(\ref{eq:RPASX_dm}) is similar to 
\ref{eq:ecorr_acfdt2b}, with 
$\b{Q}_{\alpha }^{\text{dRPA}}$ instead of 
$\b{Q}_{\alpha }^{\text{RPAx}}$ and a factor 
$1/2$ instead of $1/4$.

%The choice of the prefactor is crucial. While 
%it is clear that the single-bar integral list is always 
%accompanied by a prefactor of 1/2, the double-bar integrals 
%should be used with a prefactor of 1/4 in conjunction with the 
%("almost" correctly antisymmetrized) RPAx correlation density 
%matrix. However, in the present combination the density matrix 
%is obviously \text{non-antisymmetric}, a as it can be seen from
%its first order approximation, cf. Eq.(\ref{eq:1stordDMRPAx}). 
%It means that in no ways is justified to use the factor of 1/4
%and we have

The approximation 
$\b{Q}_{\alpha }^{-1} \approx 2\,\b{I} -\b{Q}_\alpha$ 
leads to the following approximate dRPA-II correlation energy
\begin{align}\label{eq:sosexdm}
  E_c^{\text{dRPA-IIa}}=
  \frac{1}{2}\int_0^1d\alpha  \,\,
  \text{tr}\left\{ \left[ \b{Q}^{\text{dRPA}}_{\alpha} -
  \b{I} \right] \,\b{B}\right\},
\end{align}
which is in close analogy (but \highlight{usually} not equal) to the SOSEX 
correlation energy in the ring-CCD formulation. The analytic 
relationship of this ``adiabatic-connection SOSEX'' 
(AC-SOSEX) variant with the original SOSEX has been 
discussed in detail in Ref.~\onlinecite{Jansen:10}.

%%%%%%%%%%%%%%%%%%%%%%%%%%%%%%%%%%%%%%%%%%%%%%%%%%%%%
\subsection{RPAx-I correlation energy}
\label{sec:rpaxI}
%%%%%%%%%%%%%%%%%%%%%%%%%%%%%%%%%%%%%%%%%%%%%%%%%%%%%

Finally, the RPAx-I correlation energy of \ref{eq:RPAXI} 
writes in terms of the eigenvectors of the RPAx polarization 
propagator
\begin{align}
 E_c^{\text{RPAx-I}}&=\frac{1}{2}
 \int _0^1 \!\! d\alpha \sum_n\text{Tr}
         \left\{
         \mathbb{V}\,
         \mathbb{C}_{\alpha,-n}^{~}
         \mathbb{C}_{\alpha,-n}^{\T} 
        -
         \mathbb{V}\,
         \mathbb{C}_{0,-n}^{~}
         \mathbb{C}_{0,-n}^{\T} 
         \right\},
\end{align}
leading to
\begin{equation}\label{eq:RPAxI}
   E_c^{\text{RPAx-I}}=
   \frac{1}{2}\int_0^1 d\alpha  \,
   \text{tr}
   \left\{ \left[ \b{Q}_{\alpha}^{\text{RPAx}} -
   \b{I} \right] \b{K}
   \right\},
\end{equation}
which has the same form than \ref{eq:Ec_singlebar} but 
with the RPAx matrix $\b{Q}_{\alpha}^{\text{RPAx}}$. This last 
variant has been discussed in detail and applied in the context 
of range-separated density-functional 
theory~\cite{Toulouse:09,Toulouse:10a,Zhu:10}.

%This expression can be regarded as a density matrix analog of
%the Szabo-Ostlund \textit{tilde} formula, which has been 
%derived by these authors from the ring-CCD equations.

%%%%%%%%%%%%%%%%%%%%%%%%%%%%%%%%%%%%%%%%%%%%%%%%%
\section{Correlation energy expressions in spatial-orbital basis for closed-shell systems}
\label{sec:spinadaptation}
%%%%%%%%%%%%%%%%%%%%%%%%%%%%%%%%%%%%%%%%%%%%%%%%%

For spin-restricted closed-shell calculations, all the matrices in the spin-orbital excitation basis occurring in the RPA equations have the following spin block structure
\begin{equation}
\b{C} = \left( 
  \begin{array}{cccc} 
   \b{C}_{\ua \ua,\ua \ua} & 
   \b{C}_{\ua \ua,\da \da} & 
   \b{0}                       & \b{0}                      \\ 
   \b{C}_{\da \da,\ua \ua} & 
   \b{C}_{\da \da,\da \da} & 
   \b{0}                       & \b{0}                      \\
   \b{0}                       & \b{0}                       &  
   \b{C}_{\ua \da,\ua \da} & \b{C}_{\ua \da,\da \ua}\\ 
   \b{0}                       & \b{0}                       & 
   \b{C}_{\da \ua,\ua \da} & \b{C}_{\da \ua,\da \ua}\\ 
  \end{array} 
  \right).
\end{equation}
This structure is a consequence of the fact that the two-electron integrals can be non-zero only for pairs of identical spins. The orthogonal transformation 
\begin{equation}
\b{U} = \frac{1}{\sqrt{2}} \left( 
   \begin{array}{cccc} 
   \b{1} &  \b{1}  & \b{0} &  \b{0} \\ 
   \b{1} & -\b{1}  & \b{0} &  \b{0} \\
   \b{0} &  \b{0}  & \b{1} &  \b{1} \\
   \b{0} &  \b{0}  & \b{1} & -\b{1} \\
   \end{array} 
   \right), 
\end{equation}
leads to a spin-adapted matrix $\b{\tilde{C}} = \b{U}^{\T} \, \b{C} \, \b{U}$, which in the case of the matrices involved in RPA simplifies into a block-diagonal form with a spin-singlet excitation block $^1\b{C}$ and three spin-triplet excitation blocks $^{3,0}\b{C}$, $^{3,1}\b{C}$, and $^{3,-1}\b{C}$
\begin{equation}
\b{\tilde{C}} = \left( 
  \begin{array}{cccc} 
   ^1\b{C} &         \b{0} & \b{0} & \b{0} \\ 
           \b{0} & ^{3,0}\b{C} & \b{0}  & \b{0}  \\
   \b{0} & \b{0} & ^{3,1}\b{C} &            \b{0} \\ 
   \b{0} & \b{0} &            \b{0} & ^{3,-1}\b{C} \\ 
  \end{array} 
  \right),
\end{equation}
with the matrix elements ($i,j$ and $a,b$ referring now to occupied and virtual spatial orbitals, respectively)
\begin{subequations}
\begin{align}
^1C_{ia,jb} & =\tfrac{1}{2}(C_{i\ua a\ua j\ua b\ua} + C_{i\ua a\ua j\da b\da} + C_{i\da a\da j\ua b\ua} + C_{i\da a\da j\da b\da}),
\\
%\begin{eqnarray}
%^{1/3,0}C_{ia,jb}=C_{i\ua a\ua j\ua b\ua} - C_{i\ua a\ua j\da b\da} + C_{i\da a\da j\ua b\ua} - C_{i\da a\da j\da b\da},
%\end{eqnarray}
%\begin{eqnarray}
%^{3,0/1}C_{ia,jb}=C_{i\ua a\ua j\ua b\ua} + C_{i\ua a\ua j\da b\da} - C_{i\da a\da j\ua b\ua} - C_{i\da a\da j\da b\da},
%\end{eqnarray}
^{3,0}C_{ia,jb} & =\tfrac{1}{2}(C_{i\ua a\ua j\ua b\ua} - C_{i\ua a\ua j\da b\da} - C_{i\da a\da j\ua b\ua} + C_{i\da a\da j\da b\da}),
\\
%%%\begin{eqnarray}
%%%^{3,1}C_{ia,jb}=C_{i\ua a\da j\ua b\da} + C_{i\ua a\da j\da b\ua} + C_{i\da a\ua j\ua b\da} + C_{i\da a\ua j\da b\ua},
%%%\end{eqnarray}
%\begin{eqnarray}
%^{3,1/3,-1}C_{ia,jb}=C_{i\ua a\da j\ua b\da} - C_{i\ua a\da j\da b\ua} + C_{i\da a\ua j\ua b\da} - C_{i\da a\ua j\da b\ua},
%\end{eqnarray}
%\begin{eqnarray}
%^{3,-1/3,1}C_{ia,jb}=C_{i\ua a\da j\ua b\da} + C_{i\ua a\da j\da b\ua} - C_{i\da a\ua j\ua b\da} - C_{i\da a\ua j\da b\ua},
%\end{eqnarray}
%%%\begin{eqnarray}
%%%^{3,-1}C_{ia,jb}=C_{i\ua a\da j\ua b\da} - C_{i\ua a\da j\da b\ua} - C_{i\da a\ua j\ua b\da} + C_{i\da a\ua j\da b\ua}.
%%%\end{eqnarray}
^{3,\pm 1}C_{ia,jb}& =\tfrac{1}{2}(C_{i\ua a\da j\ua b\da} \pm C_{i\ua a\da j\da b\ua} \pm C_{i\da a\ua j\ua b\da} + C_{i\da a\ua j\da b\ua}).
\end{align}
\end{subequations}

Let us start with dRPA. Spin-adaptation of the non-antisymmetrized integrals matrix $\b{K}$ gives only a contribution from the singlet excitations
\begin{equation}
\b{\tilde{K}} =  \left( 
   \begin{array}{cccc} 
   ^1\b{K} & \b{0} & \b{0} & \b{0} \\
     \b{0} & \b{0} & \b{0} & \b{0} \\
     \b{0} & \b{0} & \b{0} & \b{0} \\
     \b{0} & \b{0} & \b{0} & \b{0}\\
   \end{array} 
   \right), 
\end{equation}
where $^1 K_{ia,jb}=2\braket{ab}{ij}$. By \ref{eq:MdRPA}, it leads to the following spin-adaptation for the matrix $\b{M}^{\dRPA}_\alpha$
\begin{equation}
\b{\tilde{M}}^{\dRPA}_\alpha = \left( 
   \begin{array}{cccc} 
   ^1\b{M}^{\dRPA}_\alpha & \b{0} & \b{0} & \b{0} \\
     \b{0} & \bm{\varepsilon}^2 & \b{0} & \b{0} \\
     \b{0} & \b{0} & \bm{\varepsilon}^2 & \b{0} \\
     \b{0} & \b{0} & \b{0} & \bm{\varepsilon}^2\\
   \end{array} 
   \right), 
\end{equation}
where $^1\b{M}^\text{dRPA}_\alpha = \bm{\varepsilon}^{1/2}\, (\bm{\varepsilon}+2 \,\alpha \,^1\b{K})\,  \bm{\varepsilon}^{1/2}$, and $\bm{\varepsilon}$ refers now to the matrix of one-particle excitation energies indexed in spatial orbitals. By \ref{eq:QdRPA}, it gives the following spin-adaptation for the matrix $\b{Q}^{\dRPA}_\alpha$
\begin{equation}
\b{\tilde{Q}}^{\dRPA}_\alpha = \left( 
   \begin{array}{cccc} 
   ^1\b{Q}^{\dRPA}_\alpha & \b{0} & \b{0} & \b{0} \\
     \b{0} & \b{I} & \b{0} & \b{0} \\
     \b{0} & \b{0} & \b{I} & \b{0} \\
     \b{0} & \b{0} & \b{0} & \b{I}\\
   \end{array} 
   \right), 
\end{equation}
where $^1\b{Q}^{\dRPA}_\alpha=\bm{\varepsilon}^{1/2} \left(^1\b{M}^\text{dRPA}_\alpha\right)^{-1/2} \bm{\varepsilon}^{1/2}$.

Let us now consider RPAx. Spin-adaptation of the antisymmetrized integrals matrices $\b{A}^\prime$ and $\b{B}$ gives contributions from both singlet and triplet excitations
\begin{equation}
\b{\tilde{A}}^\prime =  \left( 
   \begin{array}{cccc} 
   ^1\b{A}^\prime & \b{0} & \b{0} & \b{0} \\
     \b{0} & ^3\b{A}^\prime & \b{0} & \b{0} \\
     \b{0} & \b{0} & ^3\b{A}^\prime & \b{0} \\
     \b{0} & \b{0} & \b{0} & ^3\b{A}^\prime \\
   \end{array} 
   \right),
\phantom{xxxx}
\b{\tilde{B}} =  \left( 
   \begin{array}{cccc} 
   ^1\b{B} & \b{0} & \b{0} & \b{0} \\
     \b{0} & ^3\b{B} & \b{0} & \b{0} \\
     \b{0} & \b{0} & ^3\b{B} & \b{0} \\
     \b{0} & \b{0} & \b{0} & -{^3}\b{B}\\
   \end{array} 
   \right), 
\end{equation}
where $^1 A^\prime_{ia,jb}=2\braket{ib}{aj}-\braket{ib}{ja}$, $^3 A^\prime_{ia,jb}=-\braket{ib}{ja}$, $^1 B_{ia,jb}=2\braket{ab}{ij}-\braket{ab}{ji}$, and $^3 B_{ia,jb}=-\braket{ab}{ji}$. Notice the minus sign for the last triplet block in the $\b{\tilde{B}}$ matrix which makes spin-adaptation less trivial for RPAx. By \ref{eq:MRPAx}, it leads to the following spin-adaptation for the matrix $\b{M}^{\RPAx}_\alpha$
\begin{equation}
\b{\tilde{M}}^{\RPAx}_\alpha = \left( 
   \begin{array}{cccc} 
   ^1\b{M}^{\RPAx}_\alpha & \b{0} & \b{0} & \b{0} \\
     \b{0} & ^3\b{M}^{\RPAx}_\alpha & \b{0} & \b{0} \\
     \b{0} & \b{0} & ^3\b{M}^{\RPAx}_\alpha & \b{0} \\
     \b{0} & \b{0} & \b{0} & ^3\b{N}^{\RPAx}_\alpha \\
   \end{array} 
   \right), 
\end{equation}
with the expected spin-adapted blocks 
$$^1\b{M}^\text{RPAx}_\alpha = \left(\bm{\varepsilon}+\alpha\, {^1}\b{A}^\prime-\alpha\, {^1}\b{B}\right)^{1/2} \left(\bm{\varepsilon}+\alpha\, {^1}\b{A}^\prime+\alpha\, {^1}\b{B}\right) \left(\bm{\varepsilon}+\alpha\, {^1}\b{A}^\prime-\alpha\, {^1}\b{B}\right)^{1/2}$$ 
and 
$$^3\b{M}^\text{RPAx}_\alpha = \left(\bm{\varepsilon}+\alpha\, {^3}\b{A}^\prime-\alpha\, {^3}\b{B}\right)^{1/2} \left(\bm{\varepsilon}+\alpha\, {^3}\b{A}^\prime+\alpha\, {^3}\b{B}\right) \left(\bm{\varepsilon}+\alpha\, {^3}\b{A}^\prime-\alpha\, {^3}\b{B}\right)^{1/2},$$ 
along with the less expected last triplet block with 
opposite signs for $^3\b{B}$, 
$$^3\b{N}^\text{RPAx}_\alpha = \left(\bm{\varepsilon}+\alpha\, {^3}\b{A}^\prime+\alpha\, {^3}\b{B}\right)^{1/2} \left(\bm{\varepsilon}+\alpha\, {^3}\b{A}^\prime-\alpha\, {^3}\b{B}\right) \left(\bm{\varepsilon}+\alpha\, {^3}\b{A}^\prime+\alpha\, {^3}\b{B}\right)^{1/2}.$$ 
By \ref{eq:QRPAx}, it gives the following spin-adaptation for the matrix $\b{Q}^{\RPAx}_\alpha$
\begin{equation}
\b{\tilde{Q}}^{\RPAx}_\alpha = \left( 
   \begin{array}{cccc} 
   ^1\b{Q}^{\RPAx}_\alpha & \b{0} & \b{0} & \b{0} \\
     \b{0} & ^3\b{Q}^{\RPAx}_\alpha & \b{0} & \b{0} \\
     \b{0} & \b{0} & ^3\b{Q}^{\RPAx}_\alpha & \b{0} \\
     \b{0} & \b{0} & \b{0} & \left(^3\b{Q}^{\RPAx}_\alpha\right)^{-1}\\
   \end{array} 
   \right), 
\end{equation}
with the spin-adapted blocks $^1\b{Q}^\text{RPAx}_\alpha = \left(\bm{\varepsilon}+\alpha\, {^1}\b{A}^\prime-\alpha\, {^1}\b{B}\right)^{1/2} \left({^1}\b{M}_\alpha^\RPAx \right)^{-1/2} \left(\bm{\varepsilon}+\alpha\, {^1}\b{A}^\prime-\alpha\, {^1}\b{B}\right)^{1/2}$ and $^3\b{Q}^\text{RPAx}_\alpha = \left(\bm{\varepsilon}+\alpha\, {^3}\b{A}^\prime-\alpha\, {^3}\b{B}\right)^{1/2} \left({^3}\b{M}_\alpha^\RPAx \right)^{-1/2} \left(\bm{\varepsilon}+\alpha\, {^3}\b{A}^\prime-\alpha\, {^3}\b{B}\right)^{1/2}$. The last triplet block turns out to be the inverse $\left(^3\b{Q}^{\RPAx}_\alpha\right)^{-1} = \left(\bm{\varepsilon}+\alpha\, {^3}\b{A}^\prime+\alpha\, {^3}\b{B}\right)^{1/2} \left({^3}\b{N}_\alpha^\RPAx \right)^{-1/2} \left(\bm{\varepsilon}+\alpha\, {^3}\b{A}^\prime+\alpha\, {^3}\b{B}\right)^{1/2}$ since according to \ref{eq:Qa} and \ref{eq:Qainv} one goes from $\b{Q}_\alpha$ to $\b{Q}_\alpha^{-1}$ by changing the sign of $\b{Y}_{\alpha}$ which is equivalent to changing the sign of $\b{B}$.

The spin-adapted correlation energy expressions can be easily 
obtained by using the invariance of the trace under the 
transformation $\b{C} \to \b{U}^{\T} \b{C} \b{U}$. The 
spin-adapted adiabatic-connection formula for the 
dRPA-I correlation energy is thus
\begin{equation}
  E_c^{\text{dRPA-I}}=
  \frac{1}{2}\int_0^1 d\alpha \, \text{tr}
  \left\{\left[^1\b{Q}_{\alpha }^{\text{dRPA}} -
  \b{I} \right]\, ^1\b{K}\right\},
\end{equation}
i.e.\ only singlet excitations contribute. Similarly, the 
corresponding plasmon formula contains only singlet excitation 
energies
\begin{align}\label{eq:dRPAIplasmon:sa}
   E_c^{\text{dRPA-I}} =\frac{1}{2} 
   \sum_n  \left( ^1\omega_{1,n}^{\text{dRPA}} - 
   {^1}\omega_{n}^{\text{dTDA}}
 \right).
\end{align}
The triplet term vanishes since both 
$^3\omega_{1,n}^{\text{dRPA}}$ and $^3\omega_{n}^{\text{dTDA}}$ 
are equal to the one-particle excitation energies 
$\varepsilon_a - \varepsilon_i$. Finally, the spin-adapted 
alternative plasmon formula is
\begin{align}\label{eq:dRPAI:sa}
   E_c^{\text{dRPA-I}}
          = \frac{1}{2}\text{tr}\biggl\{
  \left(^1\b{M}_1^{\text{dRPA}}\right)^{1/2} - 
  \left(\bm{\varepsilon}+{^1}\b{K}\right)
 \biggr\}.
\end{align}

Both singlet and triplet excitations contribute the RPAx-II 
correlation energy. The spin-adapted adiabatic-connection 
formula for RPAx-II is
\begin{align}\label{eq:EcRPAxIIsa}
    E_c^{\text{RPAx-II}} &=
    \frac{1}{4}\int _0^1 d\alpha \,  \text{tr}\left\{ 
    \tfrac{1}{2} \left({^1}\b{Q}_{\alpha}^{\text{RPAx}} \right)
    \left({^1}\b{A}'+{^1}\b{B}\right)+
    \tfrac{1}{2}\left({^1}\b{Q}_{\alpha}^{\text{RPAx}} 
    \right)^{-1}
    \left({^1}\b{A}'-{^1}\b{B}\right)-{^1}\b{A}'\right\}
\nonumber\\
    &+ \frac{3}{4}\int _0^1 d\alpha \,  \text{tr}\left\{ 
    \tfrac{1}{2} \left({^3}\b{Q}_{\alpha}^{\text{RPAx}}\right) 
    \left({^3}\b{A}'+{^3}\b{B}\right)+
    \tfrac{1}{2}\left({^3}\b{Q}_{\alpha}^{\text{RPAx}}
    \right)^{-1}
    \left({^3}\b{A}'-{^3}\b{B}\right)-{^3}\b{A}'\right\}.
\end{align}
The last triplet term gives a contribution identical to the 
other two triplet terms because the expression is invariant 
under the replacements $\b{Q}_{\alpha}\to \b{Q}_{\alpha}^{-1}$ 
and $\b{B}\to -\b{B}$. The spin-adaptation of the approximate 
RPAx-II correlation energy of \ref{eq:RPAxIIapprox} is
\begin{align}\label{eq:RPAx-IIa}
    E_c^{\text{RPAx-IIa}} &=
  \frac{1}{4}\int _0^1d\alpha  \,
  \text{tr}\left\{\left[^1\b{Q}_{\alpha}^{\text{RPAx}} -\b{I} 
  \right] \,{^1}\b{B}\right\}
  + \frac{2}{4}\int _0^1d\alpha  \,
  \text{tr}\left\{\left[^3\b{Q}_{\alpha}^{\text{RPAx}} -\b{I} 
  \right] \,{^3}\b{B}\right\}
\nonumber\\
  &- \frac{1}{4}\int _0^1d\alpha  \,
  \text{tr}\left\{\left[\left(^3\b{Q}_{\alpha}^{\text{RPAx}}
  \right)^{-1} -\b{I} \right] \,{^3}\b{B}\right\},
\end{align}
where now the last triplet term is not identical to the other 
two triplet terms. If we make the additional approximation 
$\left(^3\b{Q}_{\alpha}^{\text{RPAx}}\right)^{-1} 
\approx 2\b{I} - ^3\b{Q}_{\alpha}^{\text{RPAx}}$, 
we arrive at the following expression
\begin{align}\label{eq:RPAx-IIb}
    E_c^{\text{RPAx-IIb}} =
  \frac{1}{4}\int _0^1d\alpha  \,
  \text{tr}\left\{\left[^1\b{Q}_{\alpha}^{\text{RPAx}} -\b{I} 
  \right] \,{^1}\b{B}\right\}
  + \frac{3}{4}\int _0^1d\alpha  \,
  \text{tr}\left\{\left[^3\b{Q}_{\alpha}^{\text{RPAx}} -\b{I} 
  \right] \,{^3}\b{B}\right\},
\end{align}
which could also have been obtained by starting from the 
spin-adapted formula of \ref{eq:EcRPAxIIsa} and making the 
approximation 
$\b{Q}_{\alpha}^{-1} \approx 2\b{I} - \b{Q}_{\alpha}$ 
in both the singlet and the triplet terms. The RPAx-II plasmon 
formula decomposes into sums over singlet and triplet excitation 
energies
\begin{align}
   E_c^{\text{RPAx-II}}
     = \frac{1}{4}
         \sum_n
   \left( ^1\omega_{1,n}^{\text{RPAx}} -   
   {^1}\omega_{n}^{\text{TDAx}} \right)
   +\frac{3}{4} \sum_n \left( ^3\omega_{1,n}^{\text{RPAx}} - 
   {^3}\omega_{n}^{\text{TDAx}} \right),
\end{align}
and similarly for the alternative plasmon formula
\begin{align}
   E_c^{\text{RPAx-II}}
          = \frac{1}{4}\text{tr}\biggl\{
 \left(^1\b{M}_1^{\text{RPAx}}\right)^{1/2}
 - \left(\bm{\varepsilon}+ {^1}\b{A}^\prime\right)
 \biggr\} 
+\frac{3}{4}\text{tr}\biggl\{ \left(^3\b{M}_1^{\text{RPAx}}
\right)^{1/2} - \left(\bm{\varepsilon}+ {^3}\b{A}^\prime\right)  
\biggr\}.
\end{align}
The last triplet term is identical to the other two because 
$^3\b{N}_1^{\text{RPAx}}$ and $^3\b{M}_1^{\text{RPAx}}$ have the 
same eigenvalues and thus 
$\text{tr} \{ (^3\b{N}_1^{\text{RPAx}})^{1/2} \} = 
\text{tr} \{ (^3\b{M}_1^{\text{RPAx}})^{1/2} \}$.

The spin-adapted dRPA-II correlation energy involves only 
singlet excitations
\begin{align}
  E_c^{\text{dRPA-II}} =
  \frac{1}{2}\int_0^1 d\alpha \, 
  \text{tr}\left\{ 
  \tfrac{1}{2}
  \left({^1}\b{Q}_{\alpha}^{\text{dRPA}}\right) 
  \left({^1}\b{A}'+{^1}\b{B}\right) 
  +
  \tfrac{1}{2}
  \left({^1}\b{Q}_{\alpha }^{\text{dRPA}}\right)^{-1} 
  \left({^1}\b{A}'-{^1}\b{B}\right)-{^1}\b{A}'
  \right\},
\end{align}
since for the triplet blocks 
${^3}\b{Q}_{\alpha}^{\text{dRPA}}=\b{I}$ 
and the contribution vanishes. Likewise, the spin-adaptation of 
the approximate dRPA-II correlation energy of 
\ref{eq:sosexdm} is simply
\begin{align}
  E_c^{\text{dRPA-IIa}}=
  \frac{1}{2}\int_0^1d\alpha  \,\,
  \text{tr}\left\{ \left[ ^1\b{Q}^{\text{dRPA}}_{\alpha} -
  \b{I} \right] \, {^1}\b{B}\right\}.
\end{align}

Finally, the spin-adapted RPAx-I correlation energy 
expression is
\begin{equation}
    E_c^{\text{RPAx-I}} = 
    \frac{1}{2} \int_0^1 d\alpha  \, \text{tr} 
    \left\{ \left[ {^1}\b{Q}^{\text{RPAx}}_{\alpha} -
    \b{I} \right] \, {^1}\b{K} \right\},
\end{equation}
where only single excitations contribute since the triplet 
blocks of the matrix $\b{K}$ are zero.

%%%%%%%%%%%%%%%%%%%%%%%%%%%%%%%%%%%%%%%%%%%%%%%%%%%%%%%%%%%%%%
\section{Numerical illustrations}
\label{sec:numillust}
%%%%%%%%%%%%%%%%%%%%%%%%%%%%%%%%%%%%%%%%%%%%%%%%%%%%%%%%%%%%%%

The above-described spin-adapted RPA correlation energy variants based on the adiabatic-connection formula have been implemented in the development version of the MOLPRO quantum chemistry package~\cite{Molproshort:10}. \highlight{The numerical equality of the alternative but equivalent expressions has been carefully tested and has been confirmed within the usual accuracy of quantum chemical calculations.} In each case, we start by doing a usual Kohn-Sham (KS) calculation with some approximate density functional, and evaluate the RPA correlation energy with the KS orbitals. The total RPA energy is calculated as
\begin{equation}
E_\text{tot}^\RPA = E_{\text{EXX}} + E_c^\RPA,
\end{equation}
where $E_{\text{EXX}}$ is the exact exchange (EXX) energy  expression evaluated with the same KS orbitals. \highlight{This exchange energy is Hartree-Fock type, and it is not to be confused with the optimized effective potential (OEP) type 
local exchange, often denoted by the same acronym.} For comparison, we also perform range-separated calculations, in which we start from a range-separated hybrid (RSH)~\cite{Angyan:05}, using the short-range PBE exchange-correlation functional of 
Ref.~\onlinecite{Goll:06}, and add the long-range RPA correlation energy evaluated with RSH orbitals
\begin{equation}
E_\text{tot}^{\RSH+\RPA} = E_\RSH + E_c^{\lr,\RPA}.
\end{equation}
The long-range RPA correlation energy $E_c^{\lr,\RPA}$ is calculated by replacing the Coulombic two-electron integrals by the two-electron integrals with the long-range interaction $\erf(\mu r)/r$, just as in Refs.~\onlinecite{Toulouse:09,Toulouse:10a,Zhu:10}. We use a fixed value of the range-separation parameter of $\mu=0.5$ bohr$^{-1}$. 
\highlight{This value corresponds to a reasonable global 
compromise, as it has been shown previously~\cite{Gerber:05a}
by a study of thermochemical properties, and as it has been
confirmed later by using alternative criteria 
leading to similar estimates of the $\mu$ 
parameter (see, e.g.\ Ref.~\onlinecite{Fromager:07}).}
In all cases, the adiabatic-connection integration is performed by a 8-point Gauss-Legendre quadrature. 

\highlight{The RPA correlation energies are extrapolated to the complete basis set (CBS) limit by the usual $1/X^3$ 
formula~\cite{Kutzelnigg:92} for a series of Dunning basis sets.  In contrast to the usual two-point extrapolation
procedure~\cite{Halkier:98,Helgaker:97} all 
the available correlation energies calculated by at least 
triple zeta basis set are used. The single-determinant reference energies are evaluated with a large basis set so that they can be considered as converged.}

\begin{figure}
\begin{center}
  \subfigure[without range separation (LDA orbitals)]{\includegraphics[width=7cm]{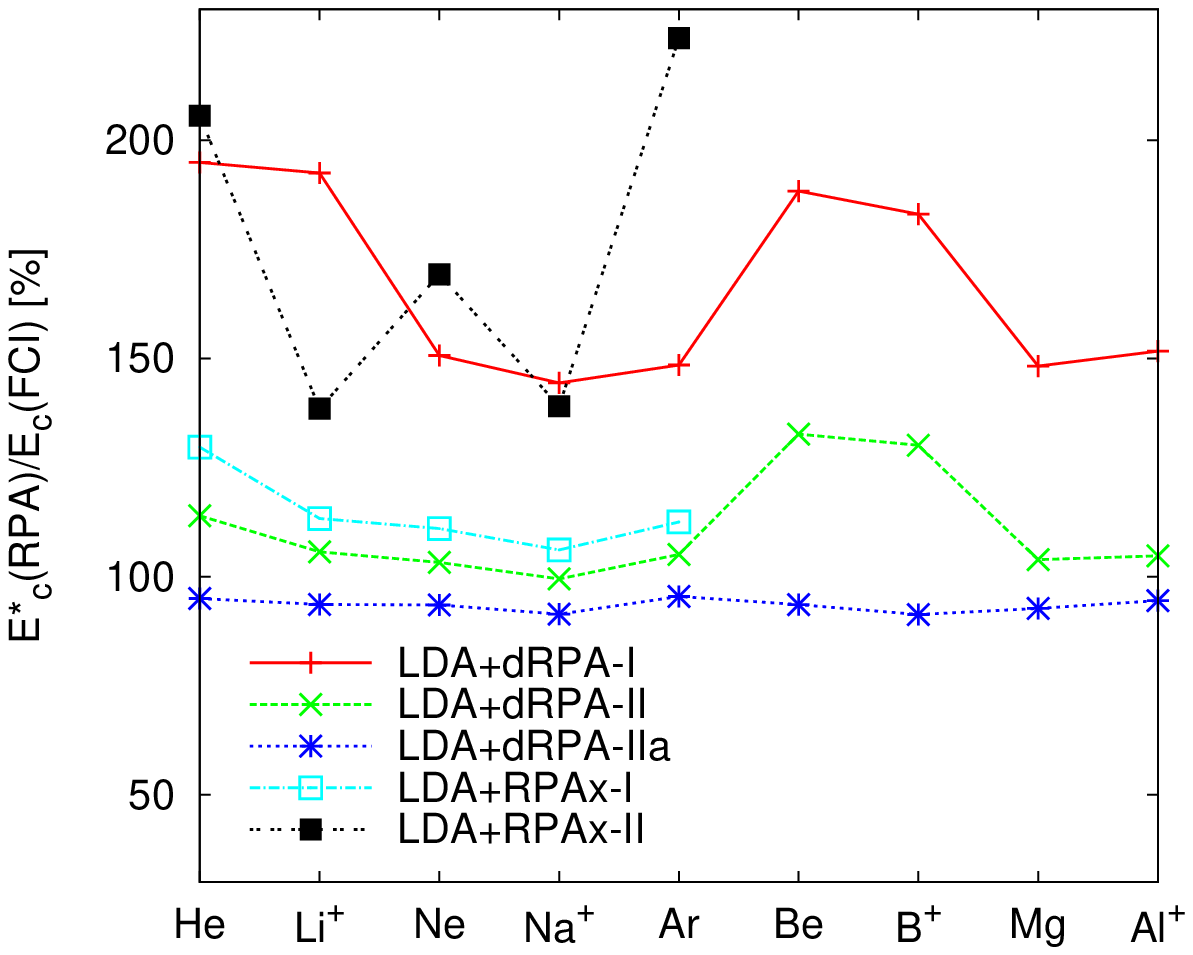}}\quad
  \subfigure[without range separation (PBE orbitals)]{\includegraphics[width=7cm]{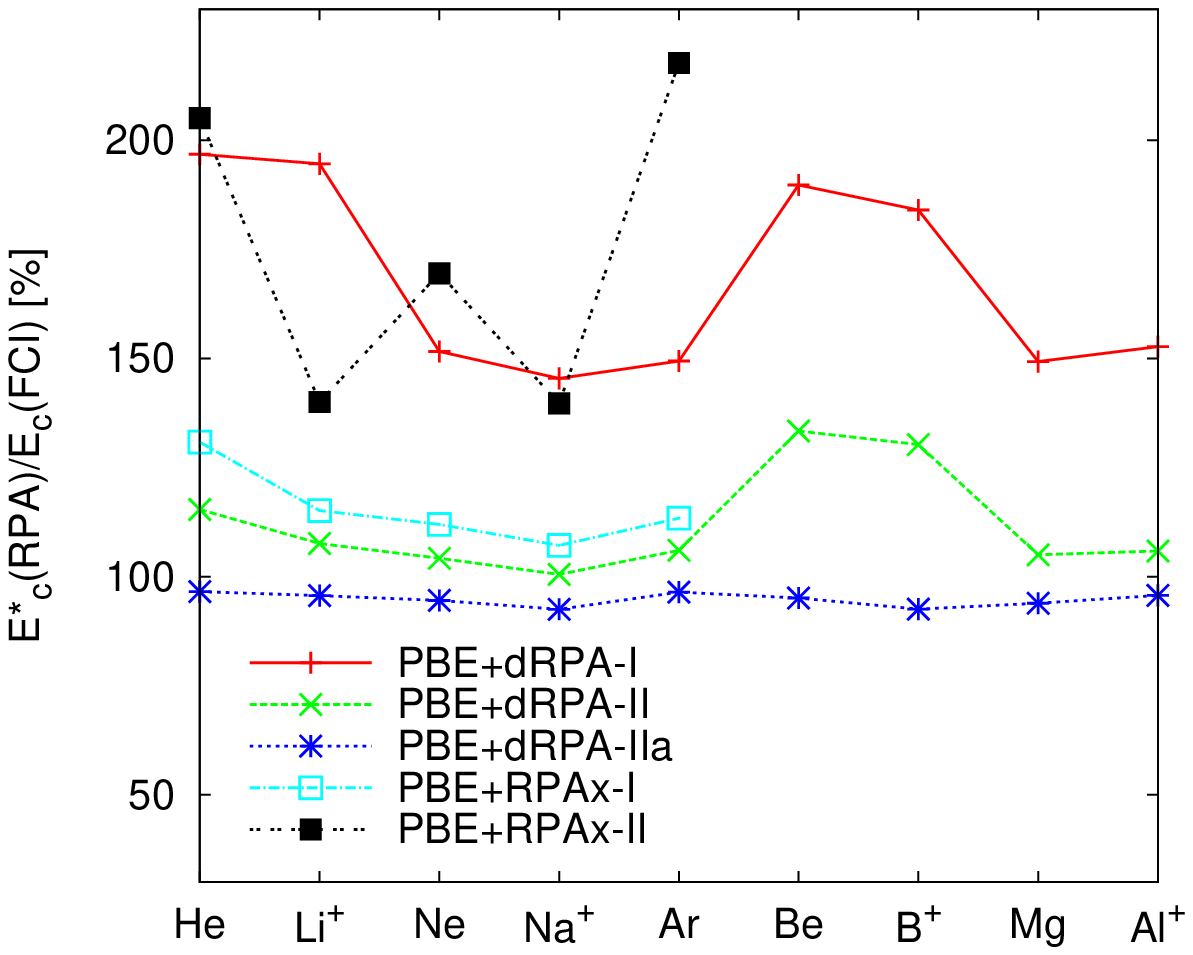}}\\
  \subfigure[without range separation (ZMP orbitals)]{\includegraphics[width=7cm]{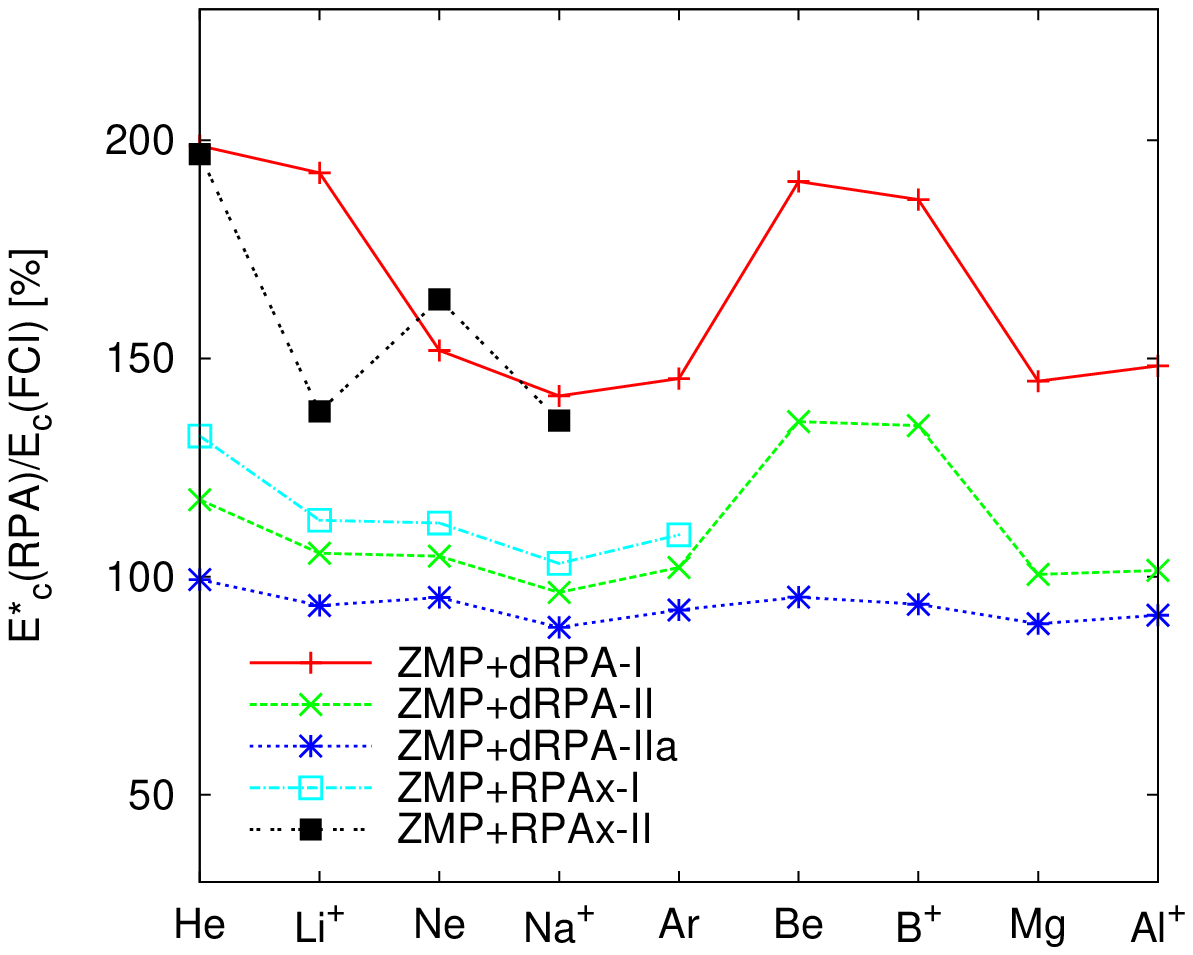}}\quad
  \subfigure[with range separation]{\includegraphics[width=7cm]{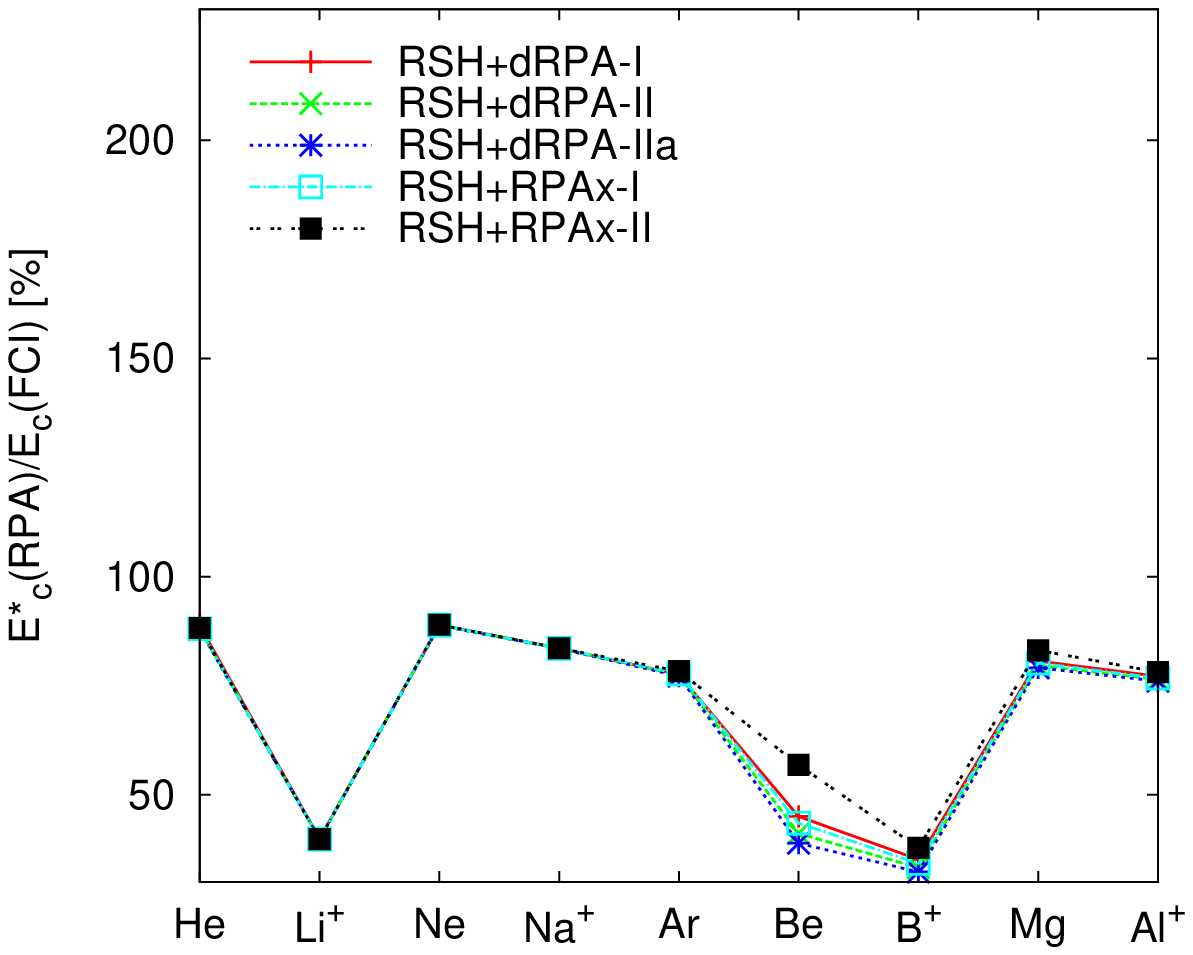}}
\end{center}
\caption{Ratios between various RPA correlation energy variants and the \highlight{FCI-quality correlation energy} as estimated by Davidson and coworkers~\cite{Davidson:91,Chakravorty:93}, with and without range separation. All the correlation energies have been extrapolated to the CBS limit. The RPA correlation energies $E_c^*(RPA)$ are redefined here as the difference between the total RPA energies and the regular HF energies.}
\label{fig:F1}
\end{figure}

\subsection{Atomic correlation energies}
\label{sec:atomicenergies}
As a first test, we have calculated correlation energies for a series of atoms and atomic cations and compared with \highlight{full configuration interaction (FCI) quality correlation energies} as estimated by Davidson and coworkers~\cite{Davidson:91,Chakravorty:93}. In order to make a direct comparison with the \highlight{FCI-quality} correlation energies which are defined with respect to the HF energies, we redefine RPA correlation energies as the difference between the total RPA energies and the regular HF energies. The single-determinant reference energies are calculated with a large even-tempered basis set. With this basis set, the HF energies agree within all significant digits with Davidson's reference data. Core excitations are included in the calculation of the RPA correlation energies and are extrapolated from the series of aug-cc-pCVXZ basis sets for He up to X=6, 
for B$^+$, Al$^+$, Ne, Ar up to X=5  
and for Li$^+$, Na$^+$, Be, Mg up to X=Q.

\ref{fig:F1}~(a)-(c) show the ratios of the correlation 
energies for each full-range RPA variant (dRPA-I, dRPA-II, 
dRPA-IIa, RPAx-I, RPAx-II) to the \highlight{FCI-quality} correlation energies, 
using orbitals obtained with the local-density approximation 
(LDA)~\cite{Vosko:80}, the Perdew-Burke-Ernzerhof 
(PBE)~\cite{Perdew:96}, and the Zhao-Morrison-Parr 
(ZMP)~\cite{Zhao:94} exchange-correlation potentials. The ZMP 
potentials have been constructed from high-quality 
\textit{ab initio} wave functions (Brueckner coupled cluster 
doubles)~\cite{Boese:privcomm}. It appears that the correlation 
energies are only marginally dependent on the quality of the KS 
orbitals, at least for this series of atomic systems. 
The full-range RPAx-I and RPAx-II variants suffer from 
instabilities in the 
RPAx response equation for the Be, B$^{+}$, Mg, and Al$^{+}$ 
systems, and additionally Ar in the case of RPAx-II with the ZMP 
orbitals. In fact, the strongly overestimated RPAx-II 
correlation energies of Ar obtained with the LDA and PBE 
orbitals indicate a situation close to an instability. More 
generally, the presence of near instabilities may be considered 
as being at the origin of the relatively strong overestimation 
(usually more than 150~\%) of the correlation energy in RPAx-II. 
In view of the poor performance of RPAx-II, we did not test the 
approximate versions of \ref{eq:RPAx-IIa} and  
\ref{eq:RPAx-IIb}. The RPAx-I variant only involves 
singlet excitations, and thus is  not subject to triplet 
instabilities. It gives quite reasonable correlation energies 
(maximum 25\% of overestimation) for He, Li$^+$, Ne, Na$^+$, and 
even for Ar.  However, RPAx-I is subject to singlet 
instabilities which appear for the rest of the systems. The 
dRPA-I variant is free of any instability problems, since the 
dRPA response matrix is positive definite by construction, but 
has nevertheless a tendency for overestimating correlation 
energies by a factor of 1.5 to 2. 
\highlight{This systematic error can be easily corrected by
including exchange in the energy expression. In fact, the} 
dRPA-II variant and  especially its approximate form dRPA-IIa
(AC-SOSEX) lead to a very good 
reproduction of the reference correlation energies.
\highlight{Similar effects could be observed recently in the 
direct ring-CCD (dRPA-I) and SOSEX calculations of 
correlation 
energies by Klopper~\emph{et~al.}~\cite{Klopper:11}, 
performed with a much smaller basis set.}  

\highlight{As mentioned previously, dRPA-IIa (or AC-SOSEX) 
and the ring-CCD-based SOSEX correlation 
energies are expected to be quite close to each other.  
Numerical results (not shown in the figures) confirm this 
expectation. 
For two-electron systems (He, Li$^{+}$) the dRPA-IIa and
SOSEX correlation energies are identical, 
while for the rest of the systems the relative
difference is less than 0.15~\%. The largest absolute 
difference, 0.82~mHartree,  has been found in full-range 
calculations on the Ne atom. It is interesting to note that 
the ring-CCD based SOSEX correlation energies are always 
lower than the dRPA-IIa values. This fact cannot be interpreted
simply by the comparison of the third order energy expressions,
reported in Ref.~\cite{Jansen:10}.}

\ref{fig:F1}~(d) shows the same total correlation energies obtained with range separation, i.e. the sum of the short-range PBE correlation energy and the long-range RPA correlation energy. The situation is quite different from the full-range RPA calculations. First, we do not encounter any instability problems anymore. Second, all the range-separated RPA variants give essentially identical correlation energies. Third, the correlation energies are systematically underestimated, for most of the systems with less than 20\% of error, but with the notable exceptions of Li$^{+}$, Be, and B$^{+}$, for which the correlation energies are underestimated by as much as 50\%. These findings may be due to the fact the systems considered here have very compact densities, and for the value of the range separation used here, $\mu=0.5$ bohr$^{-1}$, the major part of correlation is assigned to the short-range density functional rather than to the long-range RPA calculation. Improvement over these results would require either increasing the value of $\mu$ or using a more accurate short-range density-functional approximation. 

\begin{figure}
\begin{center}
  \subfigure[Bond lengths without range separation]{\includegraphics[width=7.4cm]{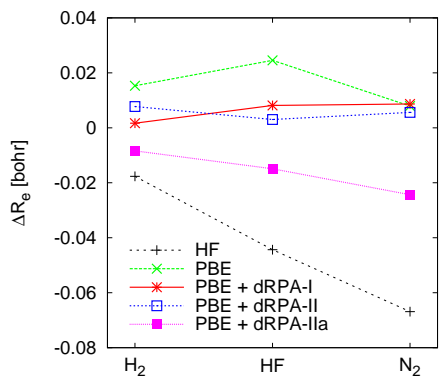}}
  \quad
  \subfigure[Bond lengths with range separation]{\includegraphics[width=7.4cm]{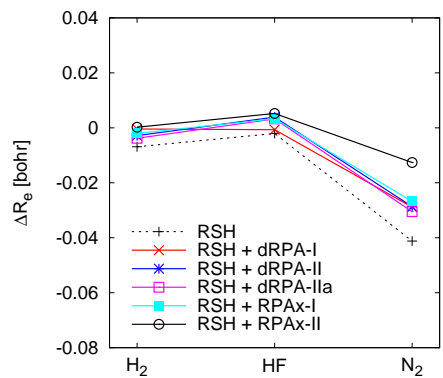}}\\
  \subfigure[Harmonic frequencies without range separation]{\includegraphics[width=7.4cm]{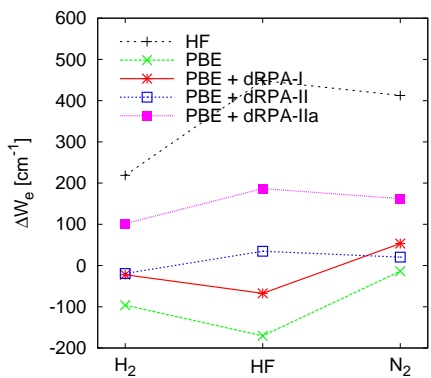}}\quad
  \subfigure[Harmonic frequencies with range separation]{\includegraphics[width=7.4cm]{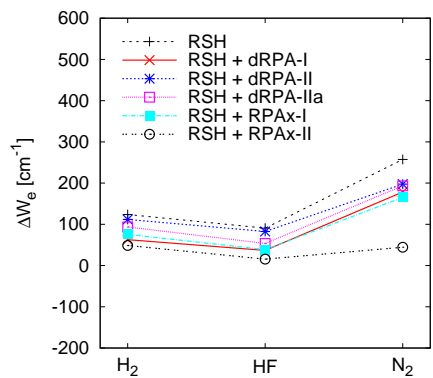}}
\end{center}
\caption{Errors in the equilibrium bond lengths and harmonic vibrational frequencies for simple diatomic molecules, calculated with the full-range and range-separated RPA variants and compared to experimental reference values. All the correlation energies have been extrapolated to the CBS limit.
The experimental reference values are (in bohr and cm$^{-1}$)
H$_2$ $R_e$ =1.40112,   $\omega_e$ = 4401.21; 
HF  $R_e$ = 1.73250,  $\omega_e$ = 4138.32;
N$_2$      $R_e$ = 2.07431,  $\omega_e$ = 2358.57 \cite{NIST:diatom}.
}
\label{fig:F2}
\end{figure}

\subsection{Bond lengths and harmonic vibrational frequencies}
\label{sec:diatomics}

\ref{fig:F2} reports equilibrium bond lengths and harmonic vibrational frequencies calculated with the full-range and range-separated RPA variants for three simple diatomic molecules, representing an apolar single bond (\ce{H2}), a strongly polar single bond (\ce{HF}), and an apolar multiple bond (\ce{N2}). The full-range RPA calculations are done with PBE orbitals, while the range-separated RPA calculations are done with the short-range PBE density functional. All RPA calculations are without core excitations, and extrapolated to the CBS limit with the series of basis sets aug-cc-pVXZ with X=T,Q,5. The single-determinant reference energies are calculated with the aug-cc-pV5Z basis set. Due to instabilities in the full-range RPAx response equation, only the full-range dRPA values can be calculated, while no instabilities are found for the range-separated RPAx calculations. Without range separation, big differences are found between the different methods. The dRPA-I and dRPA-II variants performs quite well, and represent an important improvement over both HF and KS PBE. The approximate variant dRPA-IIa is significatively less accurate than dRPA-II. With range separation, the methods give much closer results to one another. The best range-separated variant for this small set of bond lengths and harmonic frequencies appears to be RPAx-II, especially in the case
of the N$_2$ molecule.

\begin{figure}
\begin{center}
     \subfigure[He without range separation]{\includegraphics[width=7.cm]{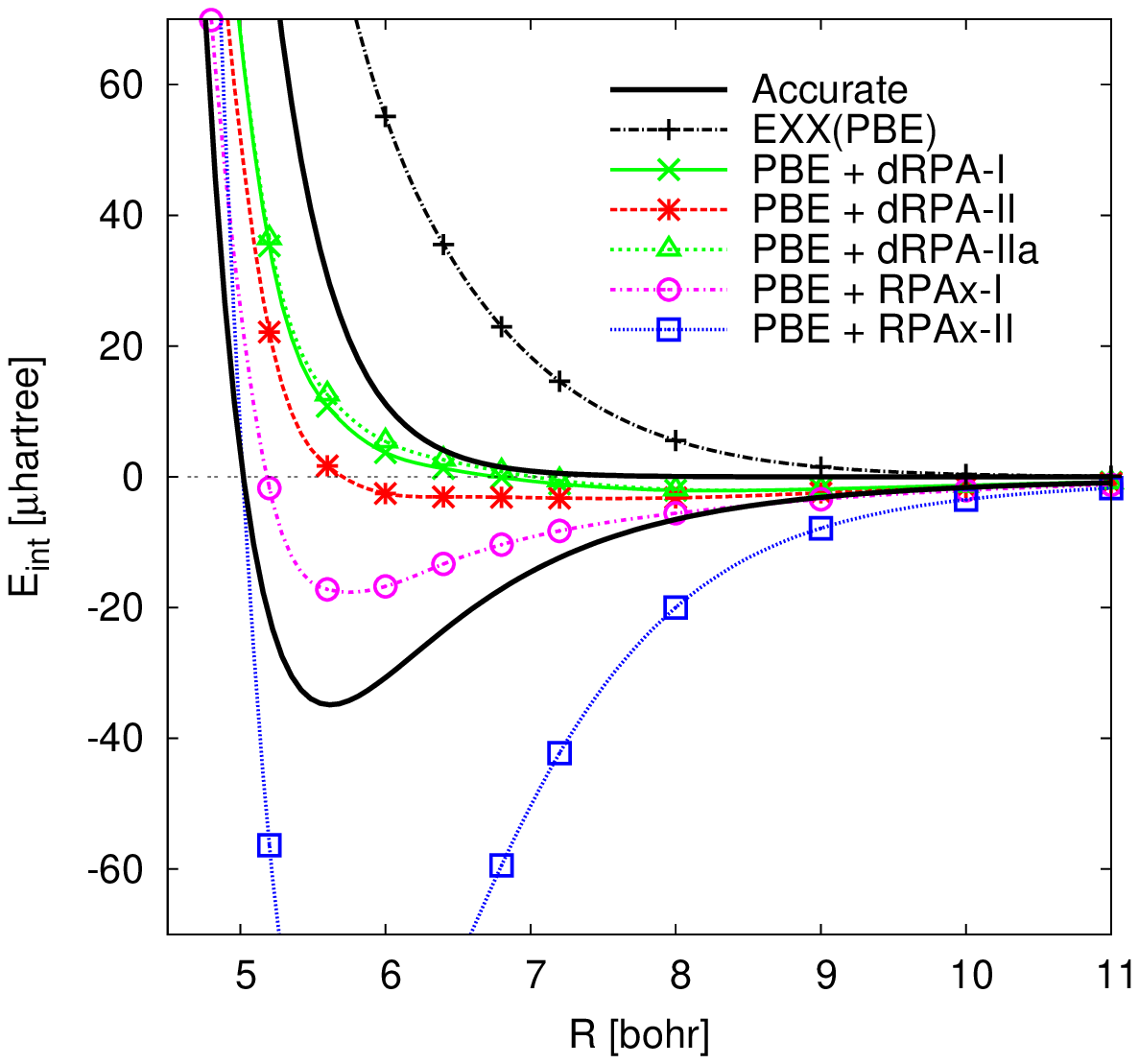}}\quad
     \subfigure[He with range separation]{\includegraphics[width=7.cm]{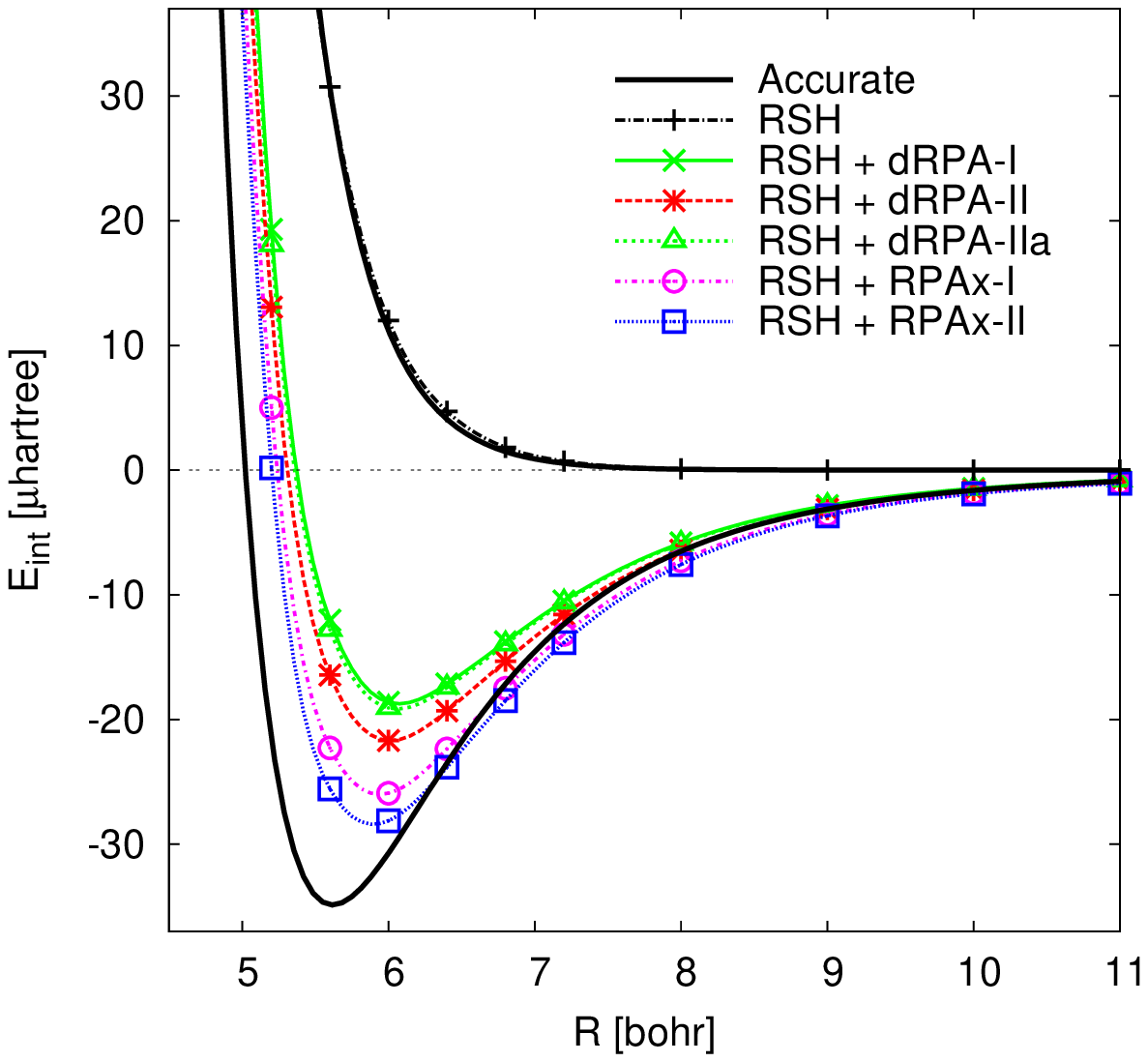}}\\\
    \subfigure[Ne without range separation]{\includegraphics[width=7.cm]{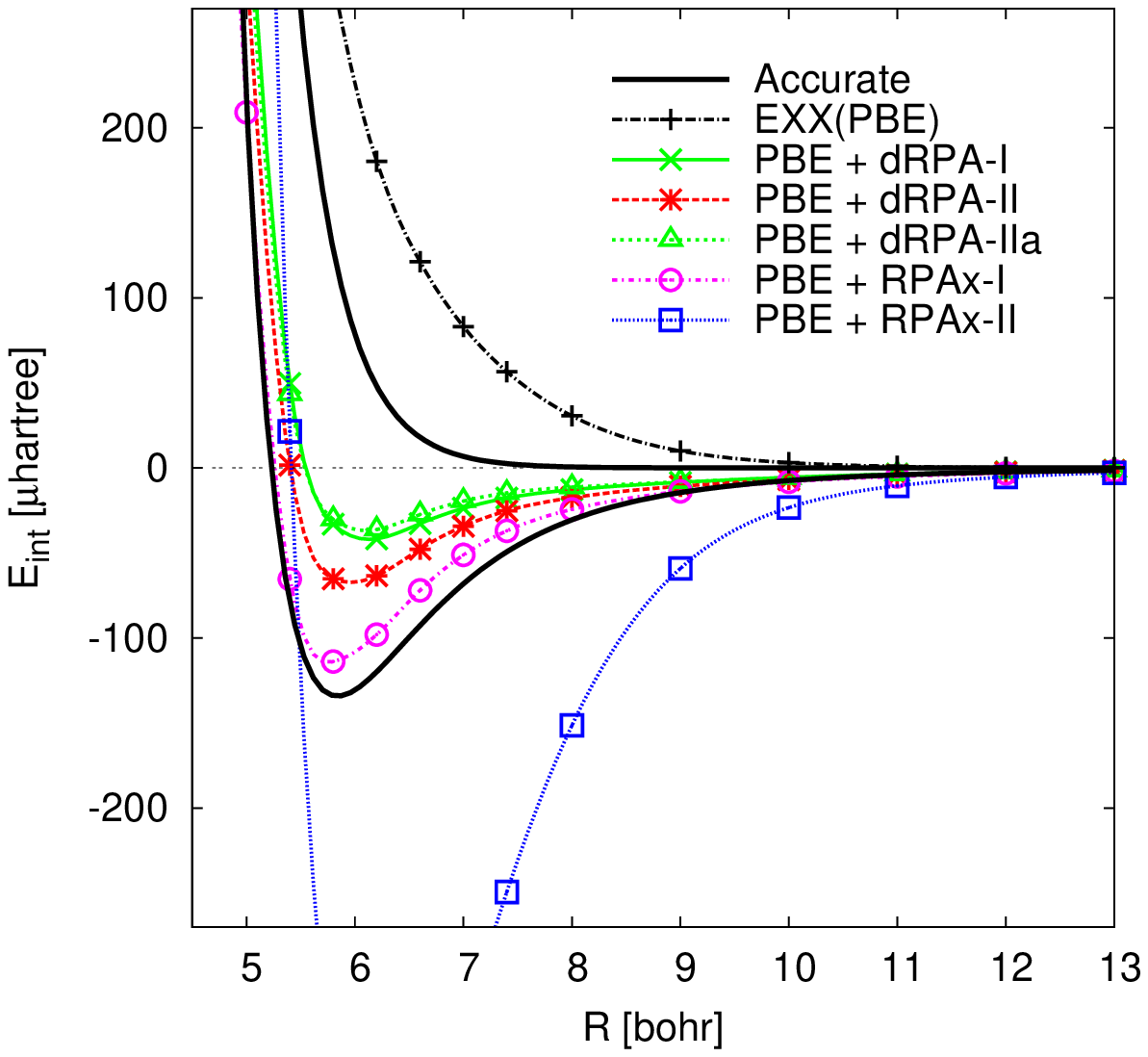}}\quad
    \subfigure[Ne with range separation]{\includegraphics[width=7.cm]{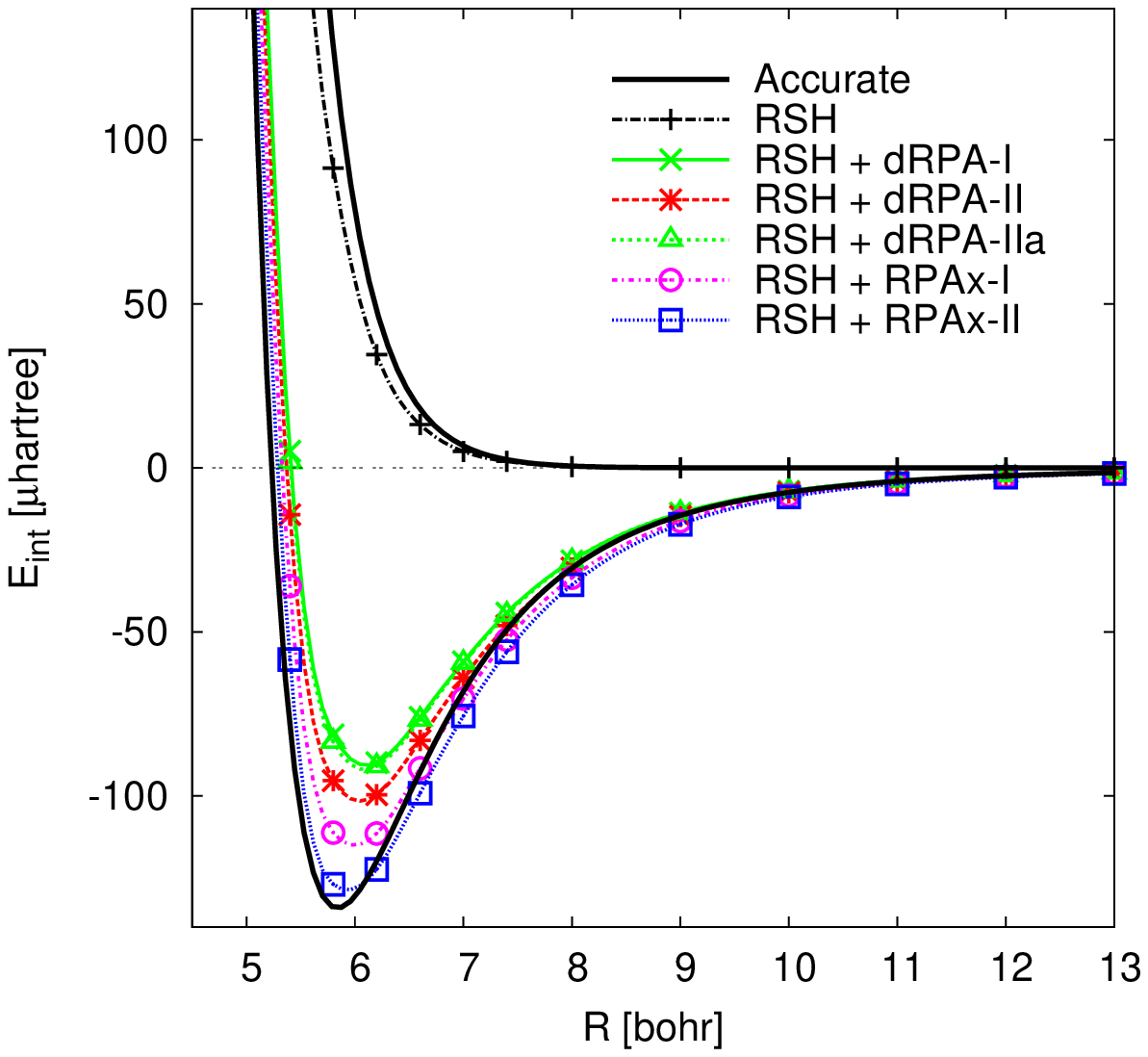}}\\  
    \subfigure[Ar without range separation]{\includegraphics[width=7.cm]{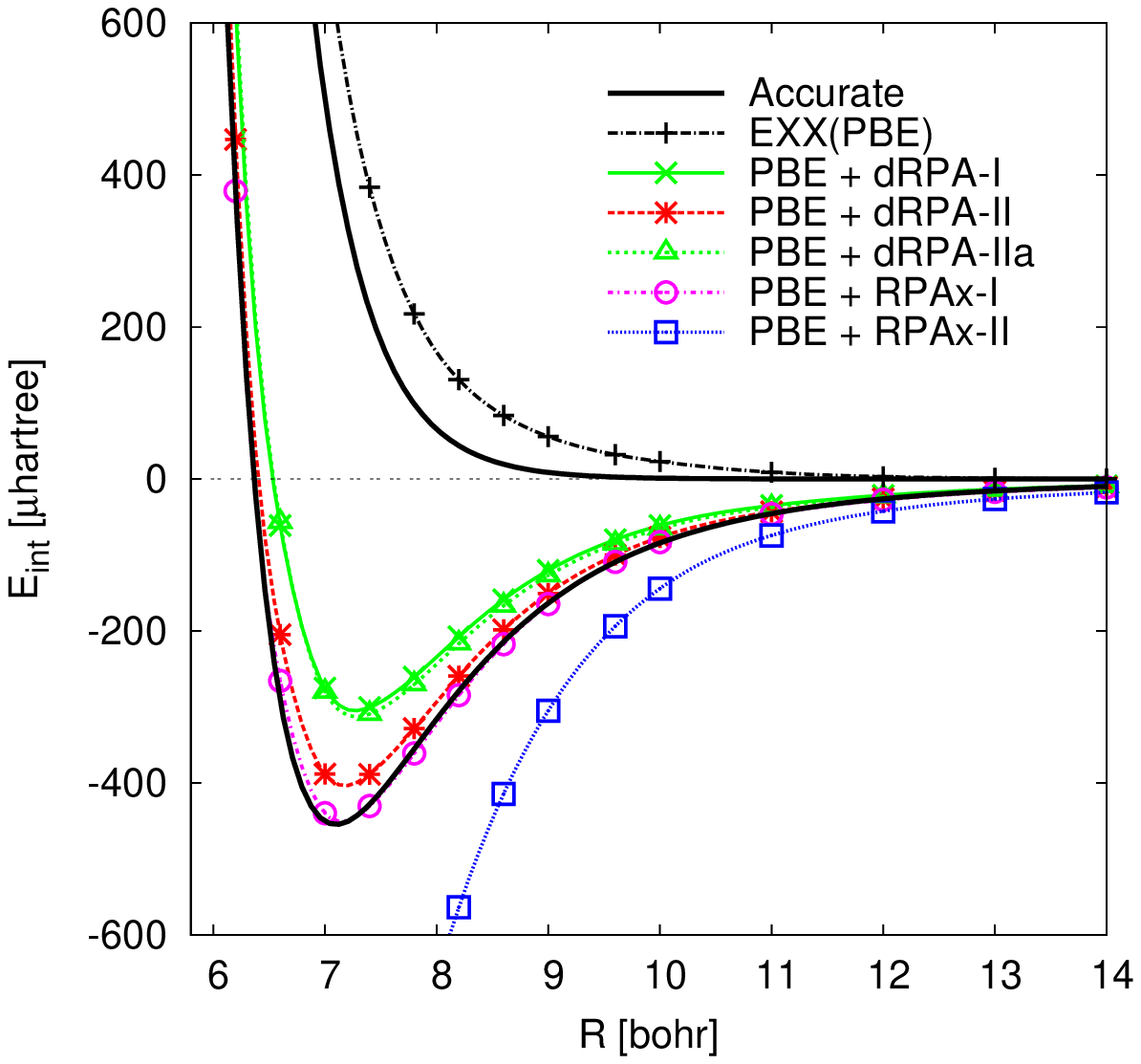}} \quad
    \subfigure[Ar with range separation]{\includegraphics[width=7.cm]{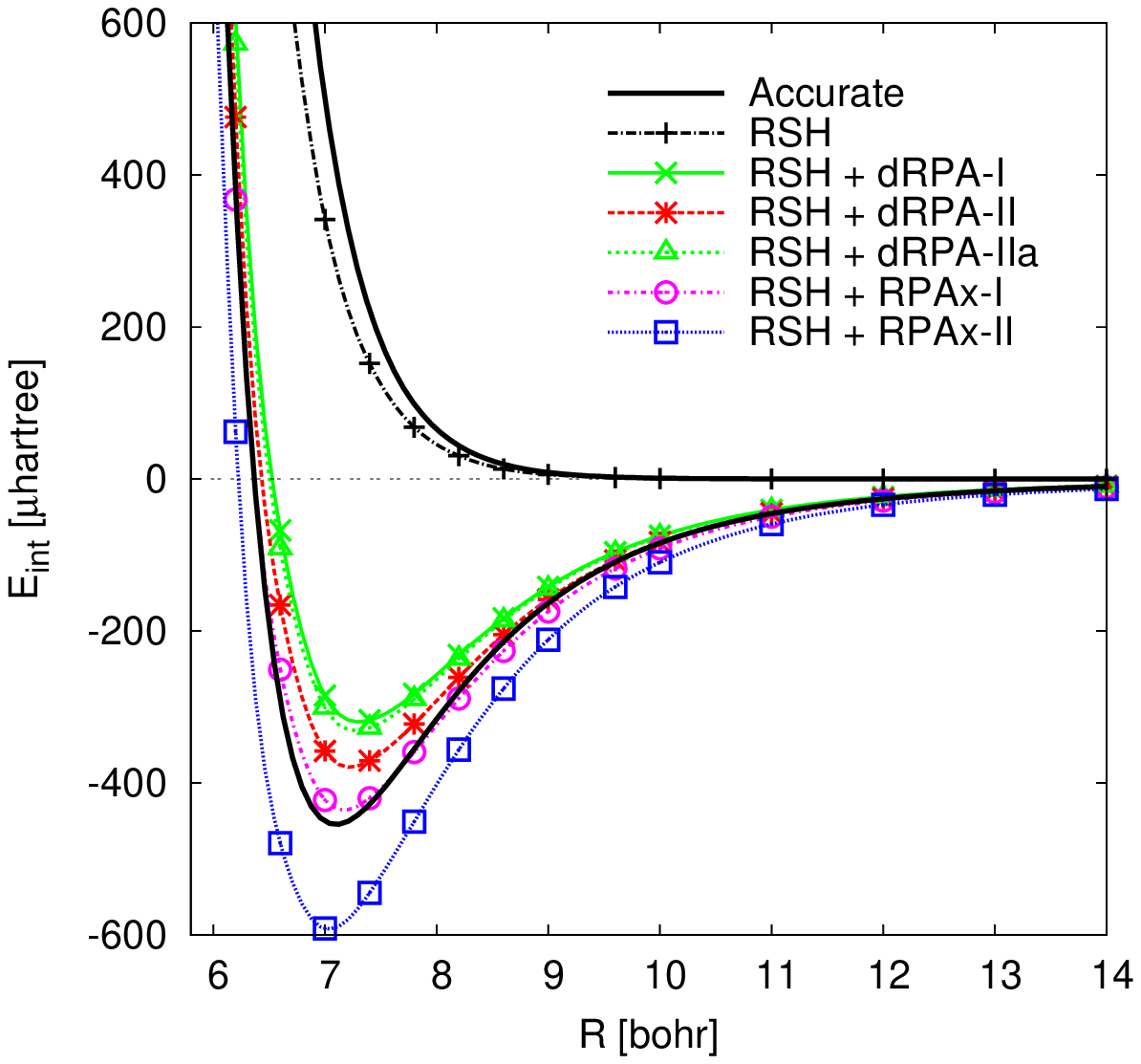}}
\end{center}
\caption{Interaction energy curves of He$_2$, Ne$_2$, and Ar$_2$, calculated with the full-range and range-separated RPA variants. All the correlation energies have been extrapolated to the CBS limit.}
\label{fig:F3}
\end{figure}

\subsection{London dispersion interactions}
\label{sec:london}

\ref{fig:F3} shows the interaction energy curves of the three rare-gas dimers He$_2$, Ne$_2$, and Ar$_2$, calculated with the full-range and range-separated RPA variants. The full-range RPA calculations are done with PBE orbitals, while the range-separated RPA calculations are done with the short-range PBE density functional. All RPA calculations are without core excitations, and extrapolated to the CBS limit with the series of basis sets aug-cc-pVXZ with X=T,Q,5,6. The single-determinant reference energies are obtained with the aug-cc-pV6Z basis set. We note that when using LDA orbitals (not shown), instabilities are found for Ne$_2$ and Ar$_2$ in a wide range of interatomic distances. In contrast, no instabilities are encountered in the case of PBE, neither with nor without range separation. 

The continuous curves without points represent on the one hand the accurate reference curves according the analytical potential energy expression of Tang and Toennies~\cite{Tang:03}, and on the other hand the repulsive (exponential) component of the same potential. These latter curves serve as  useful guides to estimate the accuracy of the single-determinant reference energies, i.e. EXX energies with PBE orbitals or RSH energies. It is quite clear that the quality of the results depends strongly on the quality of the repulsive curve. The poorest interaction energy curves are obtained for the He$_2$ dimer without range separation, for which the EXX energy is too strongly repulsive. The prerequisite of the good performance of the range-separated calculations is obviously the excellent accuracy of the RSH energy, which, for He$_2$, is in almost perfect agreement with the reference repulsive curve.
 
The full-range RPAx-II variant overestimates systematically the binding energy by a factor of 3 or more. The dRPA-I method largely underestimates the interaction energies and for He$_2$ it does not provide any minimum at all, although the non-binding character is mostly due to the bad single-determinant energy. The dRPA-II variant systematically gives more binding than dRPA-I but  also tends to underestimate the interaction energies. The approximate dRPA-IIa variant gives results that are always very close to those of dRPA-I. This is not surprising since the dRPA-I and the dRPA-IIa methods differ only by the presence of exponentially decaying exchange integrals in the interaction matrix which become quite rapidly negligible to the interaction energy in van der Waals complexes. This behavior is analogous to that of the SOSEX method which gives dispersion interaction energies also very close to those of dRPA-I~\cite{Toulouse:XX}. The best full-range method for these rare gas dimers is RPAx-I which is in quite good agreement with the reference curves for Ne$_2$ and Ar$_2$.

With range separation, all the RPA variants give much closer interaction energy curves to each other, but the same trends are found.  Range-separated dRPA-I, dRPA-II, and dRPA-IIa methods systematically underestimate interaction energies, the range-separated RPAx-II significantly overbinds Ar$_2$, and the range-separated RPAx-I globally gives the most accurate interaction energies.

%%%%%%%%%%%%%%%%%%%%%%%%%%%%%%%%%%%%%%%%%%%%%%%%%%%%%%%%%%%%%
\section{Conclusions}
\label{sec:conclusion}
%%%%%%%%%%%%%%%%%%%%%%%%%%%%%%%%%%%%%%%%%%%%%%%%%%%%%%%%%%%%%

We have analyzed several RPA correlation energy variants based 
on the adiabatic-connection formula: dRPA-I , dRPA-II, RPAx-I, 
and RPAx-II. These variants have the generic form of an 
interaction-strength-averaged two-particle density matrix 
contracted with two-electron integrals. They differ in the way 
the exchange interactions are treated. The dRPA-I variant is 
just the usual RPA of the density-functional/material-science 
community and neglects all exchange interactions. The dRPA-II 
variant uses a density matrix without exchange but contracted 
with antisymmetrized two-electron integrals. It is original to 
this work, although it resembles the SOSEX 
method~\cite{Gruneis:09}, especially in its approximate form 
named dRPA-IIa. The RPAx-I uses a density matrix with exchange 
but contracted with non-antisymmetrized two-electron integrals. 
It has previously been discussed in the context of range 
separated density-functional 
theory~\cite{Toulouse:09,Toulouse:10a}. The RPAx-II variant uses 
a density matrix with exchange and contracted with 
antisymmetrized two-electron integrals. The RPAx-II method 
itself is obviously not 
new~\cite{McLachlan:64}, but we have derived several new
expressions for it. Contracting the density matrix with either 
non-antisymmetrized or antisymmetrized two-electron integrals is 
not equivalent because of the breaking of the antisymmetry of 
the density matrix in RPA. For the dRPA-I and RPAx-II variants, 
we have made the connection with the plasmon formulation, and 
clarify the origin of the factor of $1/4$ in the plasmon formula 
for RPAx-II instead of the factor of $1/2$ for dRPA-I. We have 
carefully studied the second-order limit in the electron-
electron interaction, and showed that all the correlation energy 
variants except for dRPA-I correctly reduce to the MP2 
correlation energy (see Appendix). Finally, we have derived the spin-adapted 
forms of all these methods for closed-shell systems, and 
implemented and tested them with and without range separation of 
the electron-electron interaction.

The numerical examples on atomic and molecular systems show that the RPAx variants without range separation frequently suffer from instabilities in the RPAx response equation which make it impossible to extract a meaningful correlation energy in these cases. However, no instabilities are encountered with range separation, and the RPAx variants can be thus safely applied. The tests performed do not allow us to identify an RPA variant which would be uniformly better than the others. Without range-separation, dRPA-II performs well for atomic correlation energies and equilibrium molecular properties, but significantly underestimates London dispersion interaction energies for which RPAx-I is more accurate. With range separation, all the RPA variants tend to give more accurate results, and they also become much more similar to each other. Range-separated RPAx-II appears as the best variant for equilibrium molecular properties and range-separated RPAx-I is the best variant for dispersion interaction energies.

We hope that the overview of the RPA correlation energy variants provided in this work will be useful for a better understanding of RPA methods and can serve as a starting point for the design of improved approximations.

\section*{Acknowledgments}
We thank A. Savin (Paris) for discussions. This work was supported by ANR (Agence National de Recherche) via contract number ANR-07-BLAN-0272 (Wademecom).
G.~J. thanks Nancy University for a visiting professorship
during which part of this work has been carried out.

\appendix

%%%%%%%%%%%%%%%%%%%%%%%%%%%%%%%%%%%%%%%%%%%%%%%%%%%%%%%%%
\section{Second-order approximations to the RPA correlation energy expressions}
\label{sec:appendix}
%%%%%%%%%%%%%%%%%%%%%%%%%%%%%%%%%%%%%%%%%%%%%%%%%%%%%%%%%

In this appendix, we explicitly derive the approximations at second order in the electron-electron interaction of the RPA correlation energy variants.

We will deal with the more general RPAx response equation and obtain dRPA as a special case. We thus start from the response equation
\begin{equation}\label{RPA_equation}
(\Lambda_0 + \alpha \mathbb{W}) \mathbb{C}_{\alpha,n}
= \omega_{\alpha,n} \Delta \mathbb{C}_{\alpha,n},
\end{equation}
with 
\begin{equation}
 {\Lambda_0}=
 \begin{pmatrix}
 \bm{\varepsilon} & \b{0} \\
 \b{0} & \bm{\varepsilon}
 \end{pmatrix} ,
~~~
 {\mathbb{W}}=
 \begin{pmatrix}
 \b{A}^\prime&   \b{B}  \\
 \b{B}& \b{A}^\prime     
 \end{pmatrix} ,
~~~
 {\Delta}=
 \begin{pmatrix}
 \b{I} &  \b{0}  \\
 \b{0} & -\b{I}
 \end{pmatrix},
\end{equation}
where $\bm{\varepsilon}$ is a diagonal matrix composed of orbital energy differences $\varepsilon_{ia} = \varepsilon_a - \varepsilon_i$, and $\b{A}^\prime$ and $\b{B}$ are matrices composed of the the antisymmetrized two-electron integrals $A^\prime_{ia,jb} =\bra{ib} \ket{aj}$ and $B_{ia,jb} =\bra{ab} \ket{ij}$, and $\b{I}$ is the identity matrix.
We assume that all occupied (denoted by $i$ and $j$) and all virtual ($a$ and $b$) orbitals are real.
In the following the index pairs $ia$ and $jb$ will be replaced with simple indices $m$ and $n$. Note that the matrices are symmetric:
$A^\prime_{n,m} = A^\prime_{m,n}$ and $B_{n,m} = B_{m,n}$. 
The solutions of \ref{RPA_equation} come in pairs, {\it i.e.}, if $\mathbb{C}_{\alpha,n} = \left( \b{x}_{\alpha,n}, \b{y}_{\alpha,n} \right)$ is an eigenvector with a positive eigenvalue $\omega_{n,\alpha} > 0$, then $\mathbb{C}_{\alpha,-n} = \left( \b{y}_{\alpha,n}, \b{x}_{\alpha,n}\right)$ is an eigenvector with the negative eigenvalue $\omega_{\alpha,-n}  = - \omega_{\alpha,n}$.
In the following we will use positive integer indices to denote 
solutions which connect to positive eigenvalues in the limit 
of a vanishing coupling parameter $\alpha$,  i.e., to 
$\omega_{0,n} > 0$. Note that we also
suppose a non-vanishing HOMO-LUMO gap.

The positive energy solutions of \ref{RPA_equation} 
for $\alpha = 0$
are trivially given by $\omega_{0,n} = \varepsilon_n$,
$\b{x}_{0,n} = \b{1}_n$ and $\b{y}_{0,n} = \b{0}$,
where $\b{1}_n$ denotes the $n$-th unit
vector, i.e., a vector with vanishing components 
except for the $n$-th component which is equal to one. 
We now wish to find the first-order correction
$\mathbb{C}_{n}^{(1)}$ to the eigenvector employing 
the power-series Ansatz
\begin{align}
\omega_{\alpha,n} & =  \omega_{0,n} + 
\alpha \omega_{n}^{(1)} + \ldots  ,
\\
\mathbb{C}_{\alpha,n} & =  \mathbb{C}_{0,n} + 
\alpha \mathbb{C}_{n}^{(1)}
 + \ldots  .
\end{align}
Plugging this into \ref{RPA_equation} 
one sees that the first-order
corrections are obtained from solving
\begin{equation}\label{RPA_order1}
\Lambda_0 \mathbb{C}_{n}^{(1)} +  
\mathbb{W} \mathbb{C}_{0,n}
= \omega_{0,n} \Delta \mathbb{C}_{n}^{(1)} +
\omega_{n}^{(1)} \Delta \mathbb{C}_{0,n}  . 
\end{equation}
Multiplication of this equation from the left with 
$\mathbb{C}_{0,n}^\mathrm{T}$
and using $\mathbb{C}_{0,n}^\mathrm{T} 
\Lambda_0 \mathbb{C}_{n}^{(1)} =
\omega_{0,n} \mathbb{C}_{0,n}^\mathrm{T} \Delta
\mathbb{C}_{n}^{(1)}$
along with the normalization condition
$\mathbb{C}_{0,n}^\mathrm{T} \Delta
\mathbb{C}_{0,n} = 1$ gives the first-order correction to the eigenvalue
\begin{equation}
\omega_{n}^{(1)} = \mathbb{C}_{0,n}^\mathrm{T} \mathbb{W} 
\mathbb{C}_{0,n} = A^\prime_{n,n}.        
\end{equation}
Multiplying \ref{RPA_order1} from the left with
$\mathbb{C}_{0,m}^\mathrm{T}$ for $m \ne n$, using
$\mathbb{C}_{0,m}^\mathrm{T} \Lambda_0 
\mathbb{C}_{n}^{(1)} =
\omega_{0,m} \mathbb{C}_{0,m}^\mathrm{T} \Delta 
\mathbb{C}_{n}^{(1)}$,
and employing the orthogonalization condition
$\mathbb{C}_{0,m}^\mathrm{T} \Delta
\mathbb{C}_{0,n} = 0$ leads to
\begin{equation}
\mathbb{C}_{0,m}^\mathrm{T} \Delta
\mathbb{C}_{n}^{(1)} = - 
\frac{\mathbb{C}_{0,m}^\mathrm{T} \mathbb{W} 
\mathbb{C}_{0,n}}
{\omega_{0,m} - \omega_{0,n}}  ,
\label{C_0m}
\end{equation}
provided that the zeroth-order eigenvalues are non-degenerate, 
{\it i.e.},
that no two occupied-virtual orbital energy differences match. 
Repeating the same operations for 
$\mathbb{C}_{0,-m}^\mathrm{T}$ one arrives at
\begin{equation}
\mathbb{C}_{0,-m}^\mathrm{T} \Delta
\mathbb{C}_{n}^{(1)} =
\frac{\mathbb{C}_{0,-m}^\mathrm{T} \mathbb{W} 
\mathbb{C}_{0,n}}
{\omega_{0,m} + \omega_{0,n}}  ,
\end{equation}
where $\omega_{0,-m} = - \omega_{0,m}$ has been used. Using the resolution of identity, $\mathbb{1} = \sum_m \mathbb{C}_{0,m} \mathbb{C}_{0,m}^\mathrm{T} + \sum_{-m} \mathbb{C}_{0,-m} \mathbb{C}_{0,-m}^\mathrm{T}$, the orthogonality of $\mathbb{C}_{n}^{(1)}$ to the zeroth-order eigenvector, i.e. $\mathbb{C}_{0,n}^\mathrm{T} \Delta \mathbb{C}_{n}^{(1)}=0$, and $\Delta^2 = \mathbb{1}$, we find the expansion of the first-order correction to the positive-energy eigenvectors
\begin{equation}
\mathbb{C}_{n}^{(1)} = -\sum_{m \ne n}
\frac{\mathbb{C}_{0,m}^\mathrm{T} \mathbb{W} 
\mathbb{C}_{0,n}}
{\omega_{0,m} - \omega_{0,n}} \Delta \mathbb{C}_{0,m} + \sum_m
\frac{\mathbb{C}_{0,-m}^\mathrm{T} \mathbb{W} 
\mathbb{C}_{0,n}}
{\omega_{0,m} + \omega_{0,n}} \Delta \mathbb{C}_{0,-m} .
\label{Cvec_order1}
\end{equation}
From \ref{Cvec_order1} it follows that 
the first-order corrections read more explicitly
\begin{subequations}
\begin{align}
\b{x}_{n}^{(1)} &= -\sum_{m \ne n}
\frac{A^\prime_{m,n}}{\varepsilon_m - \varepsilon_n} \b{1}_m,
\label{X_n}
\\
\b{y}_{n}^{(1)} &= -\sum_{m}
\frac{B_{m,n}}{\varepsilon_m + \varepsilon_n} \b{1}_m.
\label{Y_n}
\end{align}
\end{subequations}
The first-order corrections to the negative-energy solutions are simply: $\omega_{-n}^{(1)} = - \omega_{n}^{(1)}$, $\b{x}_{-n}^{(1)}=\b{y}_{n}^{(1)}$, and $\b{y}_{-n}^{(1)}=\b{x}_{n}^{(1)}$.

We can obtain the first-order expansion of the matrix $\b{Q}_{\alpha}^{\text{RPAx}}$
\begin{align}
\b{Q}_{\alpha}^{\text{RPAx}} &= \sum_n (\b{x}_{\alpha,n} + \b{y}_{\alpha,n}) (\b{x}_{\alpha,n} + \b{y}_{\alpha,n})^\mathrm{T} 
\nonumber\\
&= \sum_n \b{1}_{n} \, \b{1}_{n}^\mathrm{T} + \alpha \sum_n \left[ \b{x}_{n}^{(1)} \, \b{1}_{n}^\mathrm{T} + \b{1}_{n} \, {\b{x}_{n}^{(1)}}^\mathrm{T} + \b{y}_{n}^{(1)} \, \b{1}_{n}^\mathrm{T} + \b{1}_{n} \, {\b{y}_{n}^{(1)}}^\mathrm{T} \right] + {\cal O}(\alpha^2),
\end{align}
where the sum over $n$ refers to positive-energy eigenvectors only.
The first term is simply the identity matrix
\begin{align}
\sum_n \b{1}_{n} \, \b{1}_{n}^\mathrm{T} = \b{I}.
\end{align}
Using \ref{X_n}, one can show that the term depending on $\b{x}_{n}^{(1)}$ vanishes
\begin{align}
\sum_n \b{x}_{n}^{(1)} \, \b{1}_{n}^\mathrm{T} + \b{1}_{n} \, {\b{x}_{n}^{(1)}}^\mathrm{T} &= -\sum_n \sum_{m \ne n} \frac{A^\prime_{m,n}}{\varepsilon_m - \varepsilon_n} \b{1}_m \, \b{1}_{n}^\mathrm{T} -\sum_n \sum_{m \ne n} \frac{A^\prime_{m,n}}{\varepsilon_m - \varepsilon_n} \b{1}_{n} \, \b{1}_m^\mathrm{T}
\nonumber\\
&= \b{0}.
\end{align}
This is seen by swapping $n$ and $m$ in the last term and noting that $A^\prime_{m,n}/(\varepsilon_m - \varepsilon_n)$ is antisymmetric when exchanging $m$ and $n$. Finally, using \ref{Y_n}, the term depending on $\b{y}_{n}^{(1)}$ gives
\begin{align}
\sum_n \b{y}_{n}^{(1)} \, \b{1}_{n}^\mathrm{T} + \b{1}_{n} \, {\b{y}_{n}^{(1)}}^\mathrm{T} &= -\sum_n \sum_{m} \frac{B_{m,n}}{\varepsilon_m + \varepsilon_n} \b{1}_m \, \b{1}_{n}^\mathrm{T} -\sum_n \sum_{m} \frac{B_{m,n}}{\varepsilon_m + \varepsilon_n} \b{1}_{n} \, \b{1}_m^\mathrm{T}
\nonumber\\
&= -2 \overline{\b{B}},
\end{align}
where $\overline{\b{B}}$ is the matrix with elements $\overline{B}_{m,n}=B_{m,n}/(\varepsilon_m + \varepsilon_n)$ or, more explicitly, $\overline{B}_{ia,jb}=B_{ia,jb}/(\varepsilon_a + \varepsilon_b - \varepsilon_i - \varepsilon_j)$. Therefore, we have
\begin{align}
\b{Q}_{\alpha }^{\text{RPAx}} = \b{I} - 2 \alpha \overline{\b{B}} + {\cal O}(\alpha^2),
\label{Qexpand}
\end{align}
and, similarly, the first-order expansion of the inverse matrix $(\b{Q}_{\alpha }^{\text{RPAx}})^{-1}=\sum_n (\b{x}_{\alpha,n} - \b{y}_{\alpha,n})
(\b{x}_{\alpha,n} - \b{y}_{\alpha,n})^\mathrm{T}$ yields
\begin{align}
\left(\b{Q}_{\alpha }^{\text{RPAx}}\right)^{-1} = \b{I} + 2 \alpha \overline{\b{B}} +
{\cal O}(\alpha^2).
\label{Qm1expand}
\end{align}
\ref{Qexpand} and \ref{Qm1expand} show that the approximation 
$\b{Q}_\alpha + \b{Q}_\alpha^{-1} \approx 2 \b{I}$, which lead to the definitions of $E_c^\mathrm{RPAx-IIa}$ [\ref{eq:RPAxIIapprox}] and $E_c^\mathrm{dRPA-IIa}$ [\ref{eq:sosexdm}], is correct up to first order in $\alpha$.

All the above considerations remain valid for the dRPA case, except for the replacements $\b{A}^\prime \rightarrow \b{K}$ and $\b{B} \rightarrow \b{K}$, with the obvious results
\begin{equation}
\b{Q}_{\alpha }^{\text{dRPA}}  = \b{I} - 2 \alpha \overline{\b{K}} + {\cal O}(\alpha^2)  ,
\label{QdRPAexpand}
\end{equation}
and
\begin{equation}
\left(\b{Q}_{\alpha }^{\text{dRPA}}\right)^{-1}  = \b{I} + 2 \alpha \overline{\b{K}} + {\cal O}(\alpha^2)  ,
\label{QdRPAm1expand}
\end{equation}
where the matrix elements of $\overline{\b{K}}$ are given by $\overline{K}_{m,n}=K_{m,n}/(\varepsilon_m + \varepsilon_n)$ or, more explicitly, $\overline{K}_{ia,jb}=K_{ia,jb}/(\varepsilon_a + \varepsilon_b - \varepsilon_i - \varepsilon_j)$.

We can give now the second-order limits of the RPA correlation energy variants. Using \ref{QdRPAexpand}, we find the second-order limit of the dRPA correlation energy variant of \ref{eq:Ec_singlebar}
\begin{equation}
  E_c^{\text{dRPA-I}} \approx
  \frac{1}{2}\int_0^1 d\alpha \, \text{tr}
  \left\{\left[-2 \alpha 
  \overline{\b{K}} \right]\,\b{K}\right\}
  = - \frac{1}{2} \text{tr}\left\{\overline{\b{K}}\,\b{K}\right\},
\end{equation}
which is not the normal MP2 correlation energy, but a MP2-like correlation energy without exchange, also called direct MP2 or JMP2~\cite{Janesko:09b}. In a similar way, \ref{QdRPAexpand}  and \ref{Qm1expand} give the second-order limit of the RPAx-II correlation energy variant of \ref{eq:ecorr_acfdt2b}, which is the same for its approximation of \ref{eq:RPAxIIapprox},
\begin{equation}
    E_c^{\text{RPAx-II}} \approx E_c^{\text{RPAx-IIa}} \approx
    \frac{1}{4}\int _0^1 d\alpha \, \text{tr}\left\{
    \left[-2 \alpha
    \overline{\b{B}} \right]\,\b{B}\right\}
    = - \frac{1}{4} \text{tr}\left\{\overline{\b{B}}\,\b{B}\right\},
\end{equation}
which is exactly the MP2 correlation energy expression (except for the possible replacement of Hartree-Fock orbitals and orbital energies with corresponding Kohn-Sham quantities).
The second-order limit of the dRPA-II correlation energy variant of \ref{eq:RPASX_dm} and its approximation of \ref{eq:sosexdm} are found with \ref{QdRPAexpand} and~\ref{QdRPAm1expand}
\begin{equation}
    E_c^{\text{dRPA-II}} \approx E_c^{\text{dRPA-IIa}} \approx
    \frac{1}{2}\int _0^1 d\alpha \, \text{tr}\left\{
    \left[-2 \alpha
    \overline{\b{K}} \right]\,\b{B}\right\}
    = - \frac{1}{2} \text{tr}\left\{\overline{\b{K}}\,\b{B}\right\} .
\end{equation}
Using the antisymmetry of $\b{B}$ and observing the prefactor of $1/2$, it can easily be seen that this is another way to write the usual MP2 correlation energy expression. Finally, the RPAx-I correlation energy variant of \ref{eq:RPAxI} has the following second-order limit
\begin{equation}
    E_c^{\text{RPAx-I}} \approx
    \frac{1}{2}\int _0^1 d\alpha \, \text{tr}\left\{
    \left[-2 \alpha
    \overline{\b{B}} \right]\,\b{K}\right\}
    = - \frac{1}{2} \text{tr}\left\{\overline{\b{B}}\,\b{K}\right\} ,
\end{equation}
which again exactly corresponds to the usual MP2 correlation energy expression.

Let us now consider the case of a closed-shell system. 
In this case, there is (at least) a fourfold degeneracy 
in the $\bm{\varepsilon}$ block of
$\Lambda_{0}$ since
$\varepsilon_{i\uparrow} = \varepsilon_{i\downarrow}$ and
$\varepsilon_{a\uparrow} = \varepsilon_{a\downarrow}$. 
As a consequence, the condition of non-degeneracy 
of zeroth-order excitation energies
$\omega_{0,n} = \epsilon_{ia}$ leading to 
\ref{C_0m} and \ref{X_n} is violated. Even if the final results 
for the second-order correlation energies
do not contain  differences of excitation 
energies anymore, a different 
derivation is needed. This may be achieved
by first spin-adapting the RPA response equation (for the details, see, e.g., 
Ref.~\onlinecite{Toulouse:10a}), and only subsequently making 
the perturbation expansion on the spin-adapted energy 
expressions of  \ref{sec:spinadaptation}. Assuming the absence 
of further degeneracies between 
orbital energy differences (zeroth-order excitation energies), 
one obtains formally identical expansions for the singlet and 
triplet blocks. For example, the spin-adapted 
matrices $^1\b{Q}_\alpha = \sum_n (^1\b{x}_{\alpha,n} + \, ^1\b{y}_{\alpha,n}) (^1\b{x}_{\alpha,n} + \, ^1\b{y}_{\alpha,n})^\mathrm{T}$ and 
$^3\b{Q}_\alpha = \sum_n (^3\b{x}_{\alpha,n} + \, ^3\b{y}_{\alpha,n}) (^3\b{x}_{\alpha,n} + \, ^3\b{y}_{\alpha,n})^\mathrm{T}$, where 
$(^1\b{x}_{\alpha,n},^1\b{y}_{\alpha,n})$ and 
$(^3\b{x}_{\alpha,n},^3\b{y}_{\alpha,n})$ are the singlet 
and triplet eigenvectors, and the corresponding inverse matrices  
$(^1\b{Q}_\alpha)^{-1}$ and $(^3\b{Q}_\alpha)^{-1}$ 
have the following expansions in the case of RPAx
\begin{align}
\left(^{1,3}\b{Q}_\alpha^\RPAx\right)^{\pm 1} =  \b{I} \mp 2 \alpha \, ^{1,3}\overline{\b{B}} +  {\cal O}(\alpha^2)  .
\end{align}
with $^1\overline{B}_{m,n}={^1}B_{m,n}/(\varepsilon_m + \varepsilon_n)$ and $^3\overline{B}_{m,n}={^3}B_{m,n}/(\varepsilon_m + \varepsilon_n)$. Using these results, one can easily check that all the spin-adapted correlation expressions of \ref{sec:spinadaptation} correctly reduce to MP2 at second order, except for the dRPA-I variant which reduces to direct MP2.

%%%%%%%%%%%%%%%%%%%%%%%%%%%%%%%%%%%%%%%%%%%%%%%%%%%%%%%%%%%%%%%
%% The "Acknowledgement" section can be given in all manuscript
%% classes.  Rather than use \section, an appropriate macro is
%% provided that will always work.
%%%%%%%%%%%%%%%%%%%%%%%%%%%%%%%%%%%%%%%%%%%%%%%%%%%%%%%%%%%%%%%
%\acknowledgement

%Thanks to Mats Dahlgren for version one of \textsf{achemso},
%and Donald Arseneau for the code taken from \textsf{cite} to
%move citations after punctuation.

%%%%%%%%%%%%%%%%%%%%%%%%%%%%%%%%%%%%%%%%%%%%%%%%%%%%%%%%%%%%%%%
%% The same is true for Supporting Information, 
%% which should use the \suppinfo macro.
%%%%%%%%%%%%%%%%%%%%%%%%%%%%%%%%%%%%%%%%%%%%%%%%%%%%%%%%%%%%%%%

%\suppinfo

%The entire \textsf{achemso} bundle is generated by running
%\texttt{achemso.dtx} through \TeX. Running \LaTeX\ 
%on the same file will generate the general documentation
%for both the class and package files.

%%%%%%%%%%%%%%%%%%%%%%%%%%%%%%%%%%%%%%%%%%%%%%%%%%%%%%%%%%%%%%%
%% The appropriate \bibliography command should be placed here.
%% Notice that the class file automatically 
%% sets \bibliographystyle
%% and also names the section correctly.
%%%%%%%%%%%%%%%%%%%%%%%%%%%%%%%%%%%%%%%%%%%%%%%%%%%%%%%%%%%%%%%

%\bibliography{myrefs1_iso9,myrefs2_iso9,allold_iso9,reprints_iso9}

\providecommand*\mcitethebibliography{\thebibliography}
\csname @ifundefined\endcsname{endmcitethebibliography}
  {\let\endmcitethebibliography\endthebibliography}{}

\end{document}